\documentclass{article}
\usepackage[utf8]{inputenc}
\usepackage{textcomp}
\usepackage{amsmath}
\usepackage{amssymb}
\usepackage{graphicx}
\usepackage{braket}
\usepackage{url} 
\usepackage{comment} 
\usepackage{subfigure}
\usepackage[makeroom]{cancel}
\usepackage{xcolor}
\usepackage{mathtools}
\usepackage{authblk}

  \usepackage[
    backend=biber,
    style=ieee,
  ]{biblatex}

\begin{document}

\title{\Large Madelung Mechanics and Superoscillations}

\author[1,a,*]{Mordecai Waegell}

\affil[1]{\small{Institute for Quantum Studies, Chapman University, Orange, CA 92866, USA}}

\affil[a]{ORCID: 0000-0002-1292-6041}

\affil[*]{Corresponding Author, email: waegell@chapman.edu}

\date{\today}

\maketitle

\begin{abstract}
In single-particle Madelung mechanics, the single-particle quantum state $\Psi(\vec{x},t) = R(\vec{x},t) e^{iS(\vec{x},t)/\hbar}$ is interpreted as comprising an entire conserved fluid of classical point particles, with local density $R(\vec{x},t)^2$ and local momentum $\vec{\nabla}S(\vec{x},t)$ (where $R$ and $S$ are real).  The Schr\"{o}dinger equation gives rise to the continuity equation for the fluid, and the Hamilton-Jacobi equation for particles of the fluid, which includes an additional density-dependent quantum potential energy term $Q(\vec{x},t) = -\frac{\hbar^2}{2m}\frac{\vec{\nabla}R(\vec{x},t)}{R(\vec{x},t)}$, which is all that makes the fluid behavior nonclassical.  In particular, the quantum potential can become negative and create a nonclassical boost in the kinetic energy.  This boost is related to superoscillations in the wavefunction, where the local frequency of $\Psi$ exceeds its global band limit.  Berry showed that for states of definite energy $E$, the regions of superoscillation are exactly the regions where $Q(\vec{x},t)<0$.  For energy superposition states with band-limit $E_+$, the situation is slightly more complicated, and the bound is no longer $Q(\vec{x},t)<0$.  However, the fluid model provides a definite local energy for each fluid particle which allows us to define a local band limit for superoscillation, and with this definition, all regions of superoscillation are again regions where $Q(\vec{x},t)<0$ for general superpositions.  An alternative interpretation of these quantities involving a \textit{reduced quantum potential} is reviewed and advanced, and a parallel discussion of superoscillation in this picture is given.  Detailed examples are given which illustrate the role of the quantum potential and superoscillations in a range of scenarios.

     \end{abstract}

\section{Introduction}

Superoscillation \cite{aharonov2002superoscillations,berry2006evolution, ferreira2006superoscillations, berry2008natural, aharonov2010time, aharonov2011some, aharonov2017mathematics, kempf2018four, berry2019roadmap,berry2020superoscillations,aharonov2022unified} is a wave phenomenon which occurs in both classical and quantum contexts wherein the local oscillation frequency or wave vector of a wavefunction exceeds the global band limit of that function (e.g. a wave superposition composed entirely of frequencies of smaller magnitude than $f_0$ may, in some small region, oscillate faster than $f_0$).  This phenomenon is known to occur around vortices where successive crests of the wave get squeezed together, and such vortices are known to occur around nodes in the wavefunction, and also where the wave passes through or around obstructions.

For this article, we will focus on wavefunctions $\Psi(\vec{x},t)$ satisfying the single-particle Schr\"{o}dinger equation, and their Madelung interpretation \cite{madelung1927quantum, bohm1952suggested, bohm1954model, schonberg1954hydrodynamical, holland1995quantum, wyatt2005quantum, reddiger2017madelung, reddiger2023towards, waegell2024toward}, which describes a conserved flow of a fluid in spacetime, starting from $\Psi(\vec{x},t) = R(\vec{x},t)e^{iS(\vec{x},t)/\hbar}$, where the fluid density is $\rho \equiv R^2$ and the fluid momentum is $\vec{\nabla}S(\vec{x},t) - q_0\vec{A}(\vec{x},t)$, with $q_0$ the charge of the particle, and $\vec{A}$ the vector potential.  For simplicity, we only consider the case $\vec{A}=0$ in this article.  The fluid picture is nearly classical, if one considers a new type of internal potential energy possessed by the elemental particles it comprises, called the (reduced) quantum potential energy.  It is reasonable to say that all of the apparently nonclassical behavior of this fluid is due to the role of the quantum potential energy, and if this is interpreted as a new type of classical potential energy, then the behavior of the fluid is entirely classical.  There has been some recent renewed interest in understanding the role and applications of the quantum potential \cite{sanz2019bohm,amit2023countering, shushi2023classicality, shushi2023reduced, berry2023time, berry2023quantum, silva2023properties, waegell2024toward,aharonov2024instability}.  Importantly, there is also a relatively new local quantum formalism \cite{waegell2023local} which replaces the configuration space wavefunction with a set of single-particle wavefunctions in spacetime, so even entangled many-particle systems can be understood using this fluid picture.

Michael Berry recently identified a connection between superoscillations and the quantum potential \cite{berry2020superoscillations,berry2021semiclassical}, which has already attracted some interest \cite{bloch2023spacetime}, and it is the main purpose of this article to explore and generalize these results.  Berry showed that for energy eigenstates with a single fixed value of $E$, the quantum potential $Q$ is negative in exactly the regions where the wavefunction is superoscillatory, in the sense that the local momentum of the fluid elements exceeds a corresponding `expected classical momentum,' given the energy $E$ and external potential $U$.  This also means that curves where $Q=0$ bound regions of superoscillation in the wavefunction.

While this result follows for waves of fixed energy $E$ (even superpositions with the momentum in different directions), it is no longer valid for general energy superpositions, where the quantum momentum becomes a complicated function of the different energies --- an aspect not considered by Berry.  

However, we can still consider the boundary $Q=0$ for general superpositions, and we will see that this fits with a generalization of the concept of the `expected classical energy' which incorporates all of the energies in the superposition.  We will therefore define a new class of superoscillations that are always bounded by $Q=0$.

For the analysis in this article, we work directly in terms of the energy of the fluid, rather than focusing on the momenta of elemental particles in the fluid, since the boundaries for (semiclassical \cite{berry2021semiclassical}) superoscillation can be expressed just as easily in terms of energies as in terms of momenta.  It should be stressed that this approach defines superoscillations relative to the energy eigenstates in a given superposition, and the energy eigenvalues for Hamiltonians which generally include a nontrivial external potential $U$.  This approach differs from much of the other literature on superoscillation, which considers only the Fourier components of the wavefunction, regardless of any external potential (the two cases coincide for free particles).  The notion of semiclassical superoscillation allows us to see the effect even where the wavefunction is not visibly oscillating, simply because the local momentum and kinetic energy of the particles in the fluid exceed the band limit, and this generally indicates that the fluid particles are also moving faster than the band limit would allow. 

In general, a lowercase variable represents a density and the corresponding capital variable the quantity belonging to an individual elemental particle in the fluid, and these always differ by a factor of $R^2$ (e.g., $k_a = R^2 K_a$ for average kinetic energy density).  The densities are well-behaved finite functions, which is helpful for some of the figures given in this article, although this can also obscure singularities near vortices, which are of significant interest (e.g., \cite{barnett2013superweak}), and are illustrated in other figures.

This article is organized as follows.  In Sec. \ref{Eigenstates} we review Berry's results and examine the case of real-valued energy eigenstates which solve the time-independent Schr\"{o}dinger equation, and the case of complex degenerate superpositions of such states.  In Sec. \ref{Superpositions} we examine the case of general energy superpositions, and introduce our generalized concepts of the local particle energy, the local classical kinetic energy bound, and the local bound for superoscillations.  In Sec. \ref{Reduced} we revisit and expand upon the notion of a reduced quantum potential and a symmetric kinetic energy \cite{waegell2024toward}, which is a different way to interpret the terms in the Hamilton-Jacobi equation, which leads to different concepts of the local classical kinetic energy bound, and the local bound for superoscillations.  In Sec. \ref{Examples} we go an extensive set of examples designed to build physical intuition about the fluid model and role of superoscillations, with diagrams showing various energy and energy density curves for real eigenstates and degenerate superpositions, and showing momentum streamlines for moving energy superposition for cases with both tunneling and interference.  The best visualizations for building intuition about this picture are the animations in the Supplementary Materials, with one of tunneling in a simple superposition state of a double-well potential, and two of tunneling and reflection as a Gaussian packet is incident on a finite barrier tuned as a 50/50 beam splitter.  We conclude in Sec. \ref{Conclusions} with some summarizing comments and outlook for future research.

\section{Superoscillations in Energy Eigenstates} \label{Eigenstates}

Recasting Berry's original analysis for a fixed energy $E$, the time-independent Schr\"{o}dinger equation, and a fixed potential $U(\vec{x})$, leads to the relation,
\begin{equation}
    K_a + Q = E - U,  \label{Berry}
\end{equation}
where $K_a = \frac{1}{2m}|\vec{\nabla}S(\vec{x})|^2  $ is the (average) kinetic energy, $Q \equiv  -\frac{\hbar^2}{2m}\frac{\nabla^2R(\vec{x})}{R(\vec{x})}$ is the quantum potential energy, and $K_\textrm{cl} =  E-U(\vec{x})$ is the `expected classical kinetic energy.'  The expected classical kinetic energy is negative in classically forbidden regions ($E<U(\vec{x})$), so while it follows directly from energy conservation, it is somewhat dubious to call it a classical kinetic energy.  

It is also worth noting that for stationary energy eigenstates of energy $E$, with $S=-Et$ (also called \textit{standing waves}), it follows from $K_a=0$ that the total potential $Q(\vec{x}) + U(\vec{x}) = E$ is flat (for nonzero vector potential $\vec{A}$, energy eigenstates are generally complex, and $K_a\neq 0$).  That is, the energy of every elemental particle in the fluid is $E$, and the quantum and external potential energies add up to this value everywhere in the fluid.  It then follows that the force on each elemental particle is $-\vec{\nabla}(Q+U) = 0$, which gives a classical explanation for why the fluid remains in equilibrium.  Multiplying this flatness/equilibrium condition for the fluid by $R(\vec{x})$ gives back the time-independent Schr\"{o}dinger equation for $R(\vec{x})$, 
\begin{equation}
    Q(\vec{x})R(\vec{x}) + U(\vec{x})R(\vec{x}) =-\frac{\hbar^2}{2m}\nabla^2R(\vec{x}) + U(\vec{x}) R(\vec{x}) = E R(\vec{x}),
\end{equation}
so we can think of the quantization condition as identifying the only $R_n$ functions for which $Q + U = E_n$ is flat.

Now, it is easy to see that if $K_a = E-U$, then $Q=0$, so this is the boundary of the superoscillatory region ($K_a > K_\textrm{cl} = E-U$ $\longrightarrow$ $Q<0$) for superpositions of degenerate energy eigenstates (also called \textit{stationary waves}), which necessarily includes all classically forbidden regions ($K_\textrm{cl} = E-U<0$), since $K_a \geq 0$.  Note that $K_a$ can be nonzero for degenerate superpositions of energy $E$, but the momentum $\vec{\nabla}S$ flows in closed loops such that $R^2$, $k_a$, and $q$ stay constant.  In such cases, $K_a + Q + U = E$, and thus every elemental particle in the fluid has energy $E$.  

In general, degenerate superpositions can contain nodes which produce negative singularities in the quantum potential, and the elemental fluid particles orbit in the potential wells surrounding these singularities.  Near the singularities the orbits are approximately circular, and the total force is approximately central, and thus it follows that $- \vec{\nabla} (Q+U) = \vec{\nabla} K_a = -\frac{Z}{r^3} \hat{r}$ and thus $-Q-U = K_a = \frac{Z}{2r^2} = \frac{1}{2m}|\vec{\nabla}S|^2$, where $Z$ is a constant and $r$ is the radius of the circular orbit around the singularity in $Q$ \cite{wu1994inverse}.  So we have an inverse square potential well near the singularity, with orbital speed $|\vec{v}_a| = |\vec{\nabla}S|/m = \frac{\sqrt{Z/m}}{r}$.  See the example in Sec. \ref{Vortex}.  The properties of the quantized vortices that form at nodes in the fluid density, and their relation to the Aharonov-Bohm effect, have been the studied somewhat extensively \cite{takabayasi1954hydrodynamical,takabayasi1955vector, hirschfelder1974quantized, nye1974dislocations, hirschfelder1977angular,takabayasi1983hydrodynamical,takabayasi1983vortex,wu1993quantum,reddiger2023towards,berry2023time}.

\section{Superoscillations in Superpositions of Energy Eigenstates}  \label{Superpositions}

Next we consider the corresponding relation from the time-dependent Schr\"{o}dinger equation,
\begin{equation}
    K_a + Q = -\frac{\partial S}{\partial t} - U, \label{HamJac}
\end{equation}
which has the form of a Hamilton-Jacobi equation, $-\frac{\partial S}{\partial t} = H = K_a + Q + U$, where the quantum potential energy is explicitly included.
It is easy to see that if $K_a = -\frac{\partial S}{\partial t} - U$, then $Q=0$.  For an energy eigenstate, or a superposition of terms with the same energy $E$, this reduces to Eq. \ref{Berry}, since $S(\vec{x},t)=-Et + f(\vec{x})$ for such cases.  For a general superposition, the classical energy of a particle in the fluid is,
\begin{equation}
E_p(\vec{x},t) \equiv -\frac{\partial}{\partial t}S(\vec{x},t)= K_a(\vec{x},t) + Q(\vec{x},t) + U(\vec{x},t) ,
\end{equation}
which is no longer uniform over the entire fluid.  This suggests that a natural generalization of the quantity `expected classical kinetic energy' is,
\begin{equation}
    K_\textrm{cl}(\vec{x},t) = E_p(\vec{x},t) - U(\vec{x},t) = -\frac{\partial S(\vec{x},t)}{\partial t} - U(\vec{x},t),
\end{equation}
which again makes the boundary of superoscillation $K_a > K_\textrm{cl} =  -\frac{\partial S}{\partial t} - U = K_a + Q$, or simply $Q<0$.  We can also identify the expected classical kinetic energy as equal to the \textit{quantum kinetic energy}, $K_\textrm{cl} = K_q = K_a + Q$ derived in \cite{waegell2024toward}.

If we consider cases where the external potential $U$, and thus the energy eigenvalues, are time-independent, it is natural to ask how this relates to the the maximum energy $E_+$ in a band-limited superposition of energy eigenstates.  This generalized definition will identify some regions where the local energy $K_a$ does not exceed $E_+ - U$ as superoscillatory, which may seem like a problem given the usual notion of a band limit.  However, the fluid particles cannot have energy $E_+$ everywhere, since this would make the integral over the whole fluid too large, so $E_+ - U$ is arguably too high a global bound.  The local classical energy of a fluid particle is $E_p$, and this integrates to the expectation value $\langle E \rangle$, so this generalized definition of superoscillation using $K_\textrm{cl} = E_p - U$ as the local bound seems quite reasonable, and it also reduces back to $K_\textrm{cl} = E - U$ for energy eigenstates.

That said, both bounds are still of general interest, so we define regions which exceed the local band limit but not the global band limit, $K_\textrm{cl} < K_a \leq E_+ - U$, as regions of \textit{soft superoscillation} (or simply $-\frac{\partial S}{\partial t} - E_+\leq Q<0$), and regions that exceed the global band limit, $K_a > E_+ - U$, as regions of \textit{hard superoscillation} (or $Q<-\frac{\partial S}{\partial t} - E_+$).

Note that because $K_a \geq 0$, there is always hard superoscillation in classically forbidden regions, i.e., $K_a > E_+ - U < 0$.

Finally, since the local energy of a particle in the fluid is $E_p(\vec{x},t)$ we can reasonably define $K_\textrm{cl}(\vec{x},t) <0$ as the \textit{local classically forbidden regions}, and there is always soft superoscillation in these regions, since $K_a >  K_\textrm{cl} <0$.

To summarize, by using $E_p(\vec{x},t)$ as the local energy of fluid particles, and $K_\textrm{cl}(\vec{x},t)$ as their expected kinetic energy, we obtain refined concepts of where the particles are in classically forbidden regions ($K_\textrm{cl}(\vec{x},t)<0$) and where they are superoscillatory ($K_\textrm{cl}(\vec{x},t)<K_a(\vec{x},t)$ or equivalently, $Q(\vec{x},t)<0$).  It is also noteworthy that the refined definition applies even when the external potential $U(\vec{x},t)$ varies in time, and there is no fixed energy spectrum.


\section{The Reduced Quantum Potential Energy and the Symmetric Kinetic Energy} \label{Reduced}

Now, there is a different interpretation of the quantities we have been discussing, introduced in \cite{waegell2024toward}, which breaks the quantum potential energy $Q$ apart as $Q = K_s + Q_r$, with \textit{symmetric kinetic energy} $K_s = \frac{\hbar^2}{2m} \frac{\vec{\nabla} R\cdot \vec{\nabla} R}{R^2}$ and, \textit{reduced quantum potential energy} $Q_r = -\frac{\hbar^2}{4m} \frac{\nabla^2 (R^2)}{R^2}$.  In that model, the symmetric kinetic energy $K_s$ is due to the presence of a local velocity distribution in the fluid whose symmetric part sums to zero, leaving only the average flow associated with the average kinetic energy $K_a$.  As a result of this symmetry, the total classical kinetic energy density $k_c$ of the fluid breaks apart as $k_c = k_a + k_s$.  The primary virtue of this alternative picture is that real energy eigenstates with $k_a=0$ still contain kinetic energy $k_s \neq 0$, as we would expect for the corresponding classical system, and this motion is more consistent with time-dilation in muonic atoms \cite{silverman1982relativistic}.  Note that in this picture we work in terms of the densities, because even at a given event $(\vec{x},t)$, there is a distribution of individual particle energies $K_{c_i}$, which are averaged over in the local density $k_c \equiv R^2\langle K_{c_i}\rangle_i$.
This means that the total quantum kinetic energy density can be expressed in the original Madelung/Bohm form, or in a new form using the classical kinetic energy density and the reduced quantum potential energy density, $k_q = k_a + q = k_c + q_r$.  The total energy density of elemental particles in the fluid can then be expressed as $e_p(\vec{x},t) = k_c(\vec{x},t) + q_r(\vec{x},t) + u(\vec{x}) = -R(\vec{x},t)^2\frac{\partial}{\partial t}S(\vec{x},t)$.   

The symmetric kinetic energy is suggested by the imaginary part of the quantum momentum density, and treats the energy content of each eigenstate as due to motion of the elemental particles, rather than being primarily stored in the (standard) quantum potential, with the fluid at rest.  The symmetric kinetic energy is also more directly related to the energy eigenvalue, and likewise for the momentum.
To see this, note that in 1D, $R(\pm\infty)=0$ because $R^2$ is a normalized density, and thus,
    \begin{equation}
       \langle Q_r \rangle \equiv \int_{-\infty}^\infty Q_rR^2 dx = \int_{-\infty}^\infty q_r dx =   -\frac{\hbar^2}{4m}\int_{-\infty}^\infty\frac{\partial^2(R^2)}{\partial x^2}dx 
    \end{equation}
    \begin{equation}
       = -\frac{\hbar^2}{4m}\bigg[\frac{\partial (R^2)}{\partial x}\bigg]^\infty_{-\infty} = -\frac{\hbar^2}{4m}\bigg[2R\frac{\partial R}{\partial x}\bigg]^\infty_{-\infty} = 0,   \nonumber
    \end{equation}
and this result generalizes trivially to 3D.  For real energy eigenstates, we also have $K_a = 0$, so
    \begin{equation}
        E = \langle \hat{K}\rangle + \langle \hat{U}\rangle = \langle K_a\rangle + \langle K_s\rangle + \langle Q_r\rangle + \langle U\rangle = \langle K_s\rangle + \langle U\rangle,       \nonumber
    \end{equation}
    and thus $\langle \hat{K}\rangle =  \langle K_s\rangle \equiv \int_{\textrm{space}} K_s R^2 dV$, so all of the kinetic energy in the eigenstate is apparently of the symmetric form, as promised.

The symmetric kinetic energy density arises because the elemental fluid particles have a symmetric velocity distribution $|\vec{v}_s(\vec{x},t)|\hat{r}_i(\vec{x},t)$, with $\sum_i \hat{r}_i(\vec{x},t) =0$ at each point in spacetime, and this is why it is unrelated to the average flow $\vec{v}_a(\vec{x},t)$.  Thus, relative to the average flow, the symmetric kinetic energy effectively behaves as another local potential energy, which is why the $(q,k_a)$ picture is often more intuitive than the $(q_r, k_c)$ picture.

One speculative model that would explain this is that the symmetric velocity distributions are such that the elemental fluid particles at each location $\vec{x}$ fall into a circular potential well with infinitesimal radius, where they orbit at speed $|\vec{v}_s(\vec{x},t)|$ with nearly infinite frequency.  This model is inspired by the macroscopic vortex in the 2D square well discussed in the example of Sec. \ref{Vortex}, where the average flow velocity $\vec{v}_a(\vec{x},t)$ is associated with the circuitous motion around the singularity in $Q$.  Here we would say that the portion of $Q = Q_r + K_s$ associated with $Q_r$ is a local internal energy storage mechanism, while the portion associated with $K_s$ is comprised of a sea of infinitesimal vortices where some of the kinetic energy of the particle can be trapped in the infinitesimal orbits ($K_s$), and some can be involved in net flow ($K_a$).  Finally, we have only used the symmetric speed $|\vec{v}_s(\vec{x},t)|$ in the model so far, but the bare symmetric velocity $i\vec{v}_s(\vec{x},t)$ emerges from the quantum momentum density with a direction, so we interpret it as a rotational vector, which points along the axis of rotation, and interpret the $i$ as indicating that the particle orbits the axis $\hat{v}_s(\vec{x},t)$ counterclockwise with speed $|\vec{v}_s(\vec{x},t)|$ at an infinitestimal radius.  


We now consider the circumstances of superoscillation in this alternative interpretation, where $q_r$ and $k_c$ are the physically significant quantities.  The arguments for the expected classical kinetic energy are unchanged, so we have the lower bound $k_\textrm{cl} = e_p - u = k_q = q_r + k_c$ for soft superoscillation, and the lower bound $R^2E_+ - u$ for hard superoscillation.  However, in this picture, we must check the total classical kinetic energy density $k_c$ for superoscillation, rather than only the average kinetic energy $k_a$ (which is zero for stationary states anyway), and thus we define regions of soft superoscillation to be where $k_c > k_\textrm{cl} = q_r + k_c$, or $q_r<0$, so $q_r=0$ defines the boundary for soft superoscillation in the $(q_r, k_c)$ picture, just as $q$ does in the original $(q, k_a)$ picture.  The regions of soft superoscillation give way to regions of hard superoscillation where $k_c > R^2E_+ - u$.   Hard superoscillation occurs where $q_r<-R^2\big(\frac{\partial S}{\partial t} +E_+\big)$.

Note that because $k_c \geq 0$, there is always hard superoscillation in classically forbidden regions, i.e., $k_c > R^2E_+ - u < 0$, and soft superoscillation in the local classically forbidden regions regions, since $k_c(\vec{x},t) >  k_\textrm{cl} <0$, analogous to the $(q,k_a)$ picture.

Finally, because $k_c \geq k_a$, the region of superoscillation is generally larger in the $(q_r,k_c)$ picture than in the $(q, k_a)$ picture, which can be seen throughout Sec. \ref{Examples}.

\section{Examples} \label{Examples}

It is very helpful to go through several key examples to build up a physical intuition for how things work in the fluid picture, and under what circumstances we will find superoscillations.

\subsection{Energy Superposition in the Infinite Square Well}

\begin{figure}
\centering
\begin{tabular}{c}
     \includegraphics[width=3.75in]{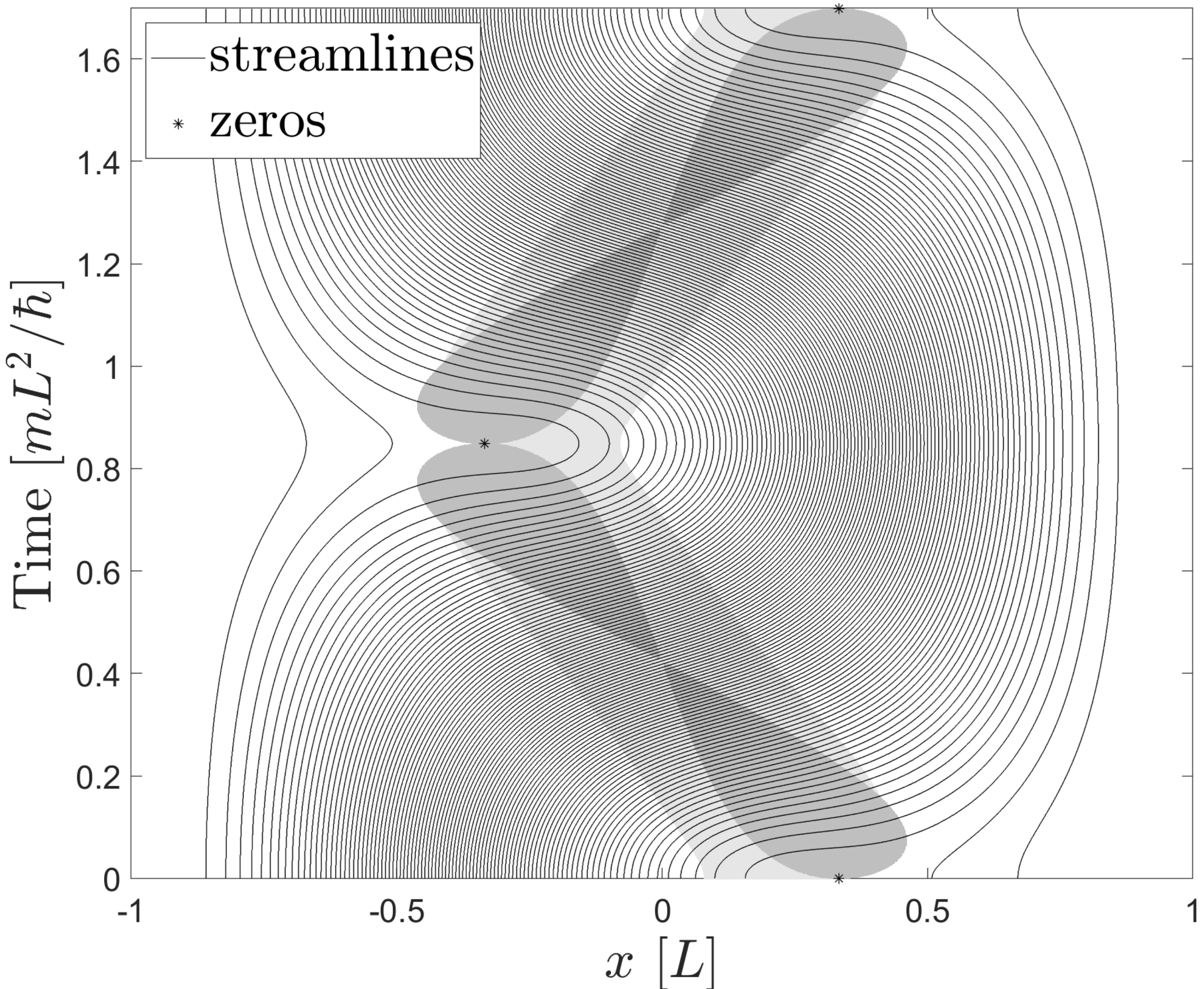}\\\\ \includegraphics[width=3.75in]{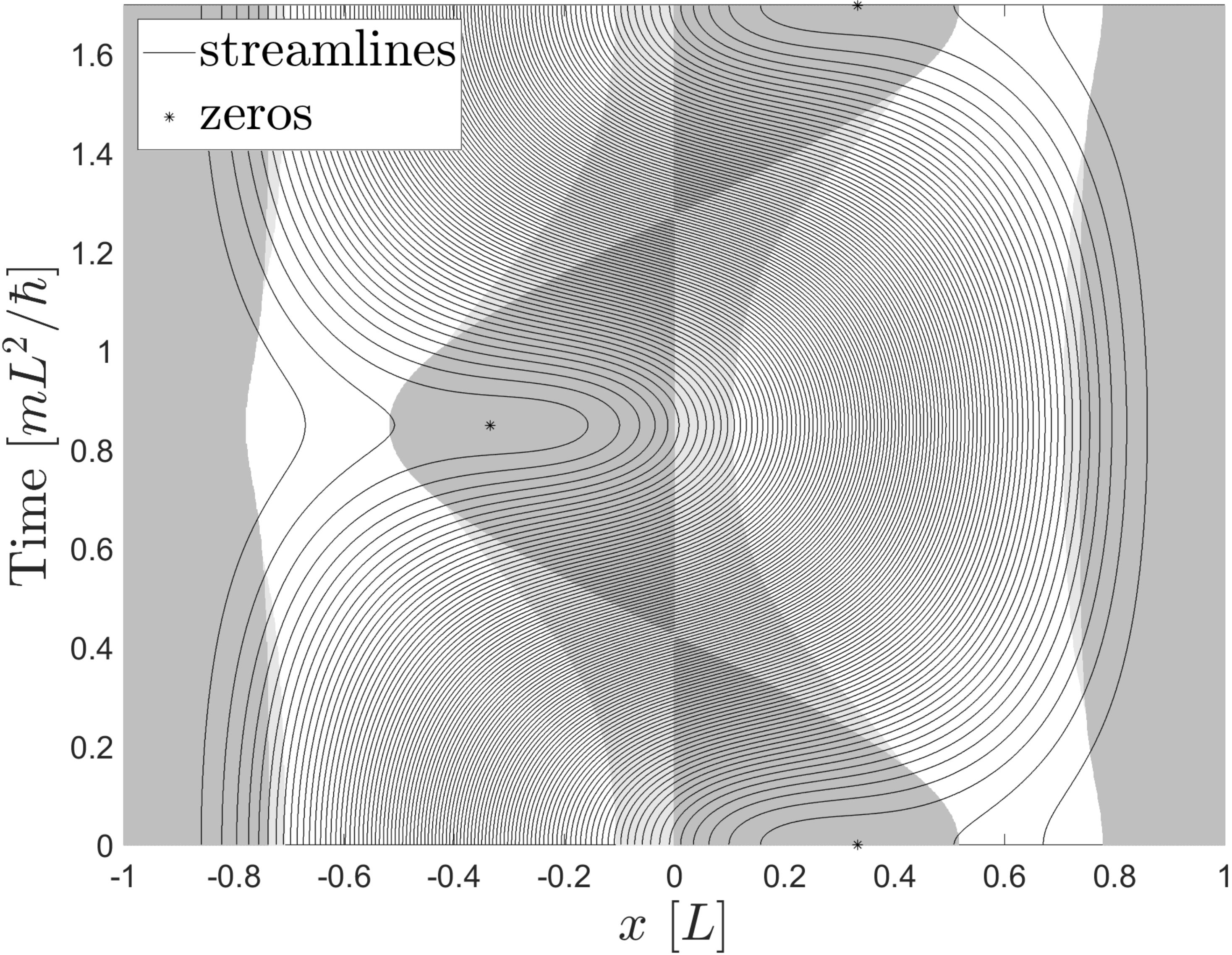}
\end{tabular}
    \caption{Streamline diagrams for one period of evolution for an equal superposition of the ground state and first excited state of a 1D infinite square well of width $2L$.  Neighboring streamlines are separated by equal proportions of fluid, and so they depict the fluid density.  The nodes that appear once per cycle are shown with asterisks.  \textbf{upper:} The lighter gray shading indicates regions of soft superoscillation, $Q<0$ and the darker gray shading indicates regions of hard superoscillation $K_a > 4E_1$ ($U=0$ inside the well).  \textbf{lower:} The reduced quantum potential picture.  The darker gray shading indicates regions of soft superoscillation, $Q_r<0$ and the darker gray shading indicates regions of hard superoscillation $K_c > 4E_1$.}
    \label{fig:InfSqStreams}
\end{figure}

One of the simplest cases of a bound state which varies in time is an equal superposition of the ground state and the first excited state in the 1D infinite square well.  For a well with ends at $x=-L$ and $x=L$, this state is,
\begin{equation}
    \psi(x,t) =\frac{1}{\sqrt{2L}}\bigg(\cos\Big(\frac{\pi x}{2L}\Big)e^{-iE_1t/\hbar} +  \sin\Big(\frac{2\pi x}{2L}\Big)e^{-4iE_1t/\hbar} \bigg),
\end{equation}
where $E_1 = \frac{\pi^2\hbar^2}{8mL^2}$ is the ground state energy, and the first excited energy is $E_2 = 4E_1$.  These parameters were chosen for easy comparison with the next example where a finite barrier is added around $x=0$.  This state evolves periodically with angular frequency $3E_1/\hbar$, with nodes appearing on either side once every cycle.  The situation is illustrated in Fig. \ref{fig:InfSqStreams}, with streamlines indicating the motion of the fluid and shading indicating areas of superoscillation.

Now, considering the upper diagram in Fig. \ref{fig:InfSqStreams}, superoscillation appears to play a direct role in this motion, and the cyclic motion appears to be driven by a dip, where the quantum potential becomes negative, that appears to `bounce' back and forth between the two events where the nodes form.  The dip ($Q<0$, so soft superoscillation) creates a local boost in the flow kinetic energy $K_a$ as the fluid is pulled across, which is what really drives the motion of the fluid, and in some regions this is enough kinetic energy to constitute hard superoscillation.

Considering the lower diagram in Fig. \ref{fig:InfSqStreams}, in the $(Q_r, K_c)$ picture, we see the same role for the reduced quantum potential in driving the flow, but the wells are significantly deeper in some places, and some of the energy is stored locally in the form of symmetric kinetic energy $K_s$, so the regions of soft superoscillation where $Q_r<0$, or hard superoscillation where $K_c = K_a + K_s > 4E_1$ are generally larger than in the standard $(Q,K_a)$ picture.

\subsection{Tunneling Through a Finite Barrier in the Infinite Square Well}

\begin{figure}
\centering
\begin{tabular}{c}
     \includegraphics[width=3.2in]{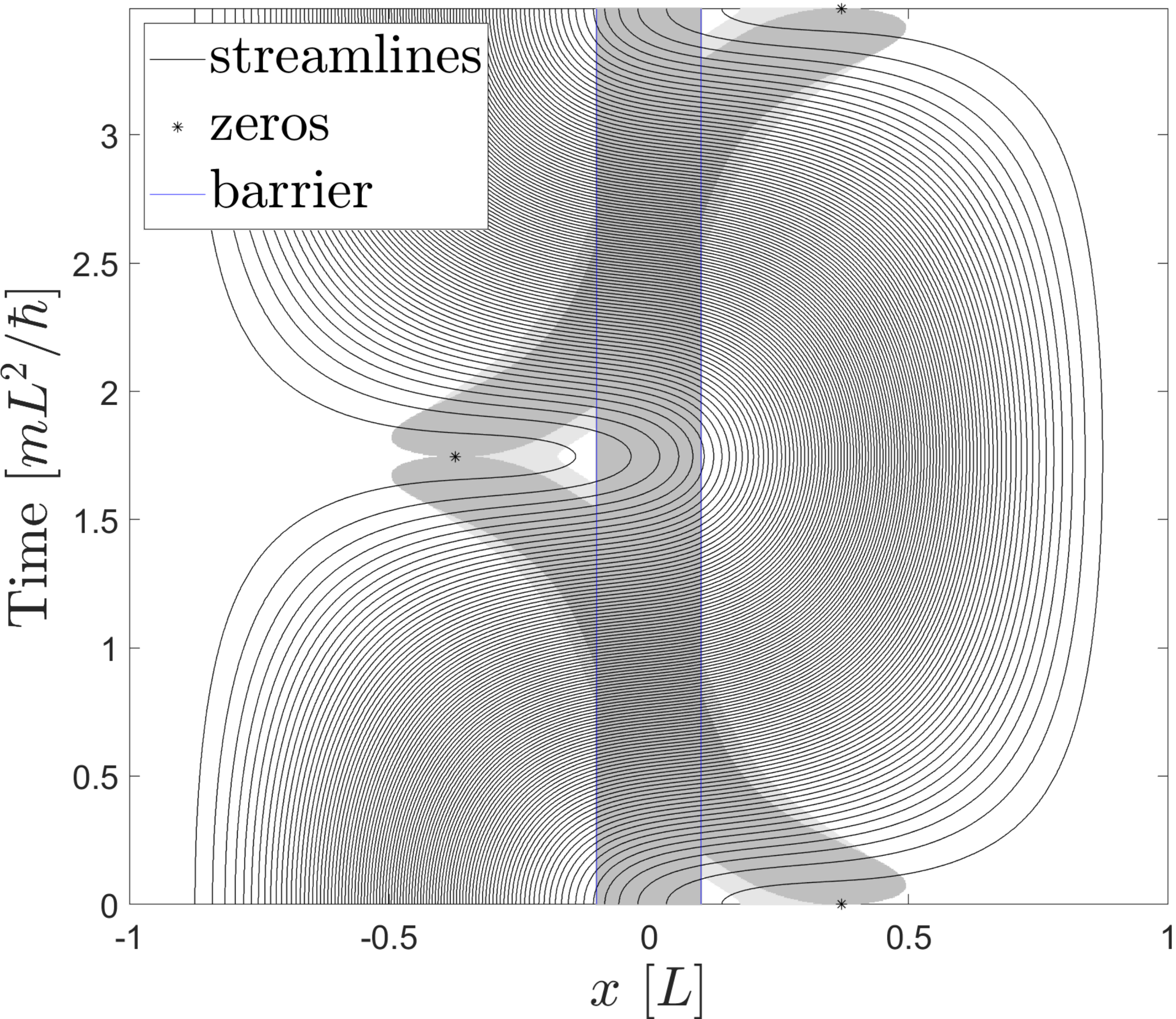}\\\\ \includegraphics[width=3.2in]{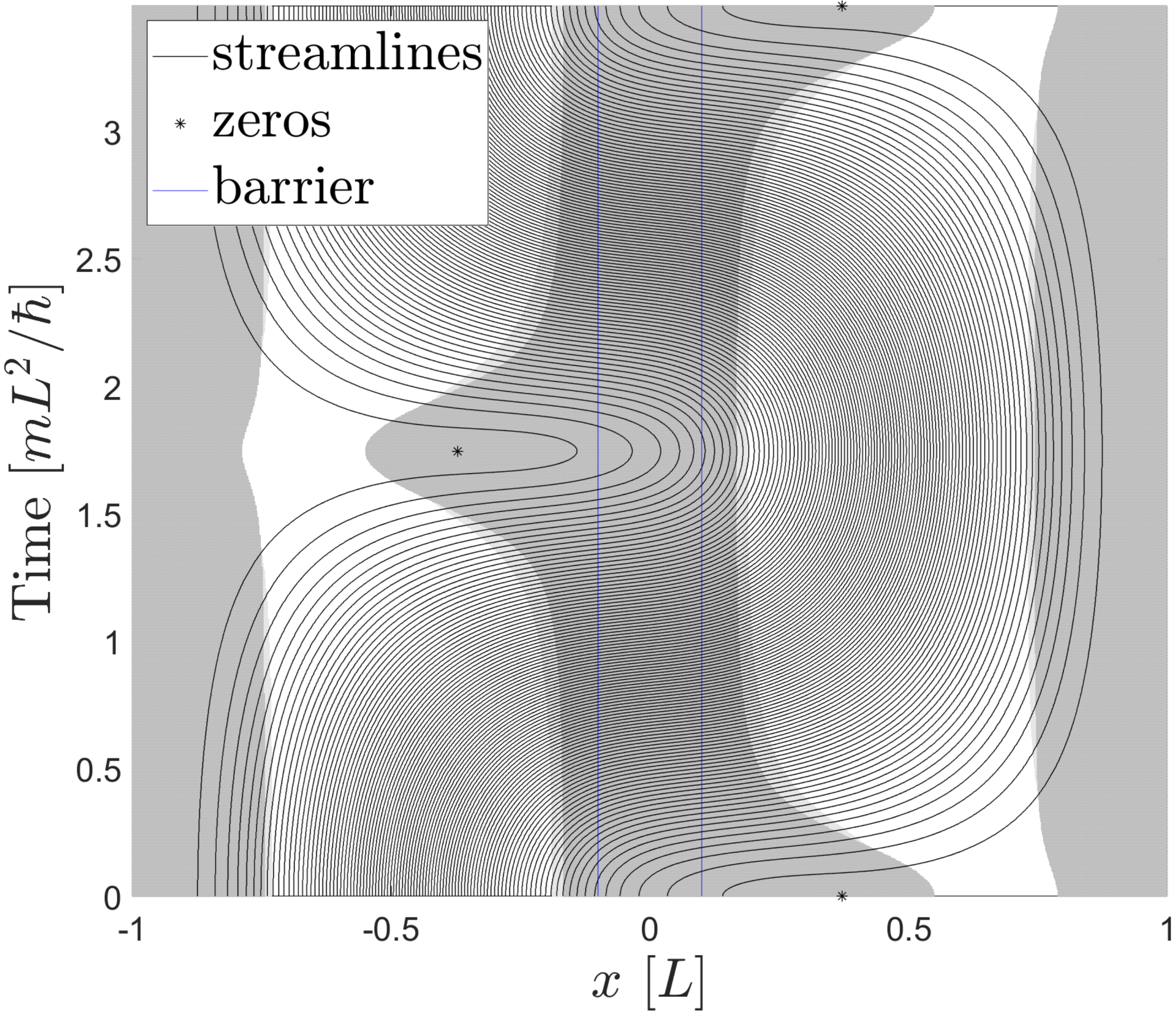}
\end{tabular}
    \caption{Streamline diagrams for one period of evolution for an equal superposition of the ground state and first excited state of a 1D infinite square well of width $2L$ with a finite barrier of height $U_0$ and width $L/5$ in the center.  Neighboring streamlines are separated by equal proportions of fluid, and so they depict the fluid density.  The nodes that appear once per cycle are shown with asterisks.  \textbf{upper:} The lighter gray shading indicates regions of soft superoscillation, $Q<0$ and the darker gray shading indicates regions of hard superoscillation $K_a > E_2'-U$.  \textbf{lower:} The reduced quantum potential picture.  The darker gray shading indicates regions of soft superoscillation, $Q_r<0$ and the darker gray shading indicates regions of hard superoscillation $K_c >  E_2'-U$.  The fluid tunnels through the barrier within continuous central region of superoscillation.}
    \label{fig:InfSqStreamsqr}
\end{figure}

We continue with the 1D infinite square well, but we have now added a finite rectangular barrier of height $U_0 = 15\hbar^2/mL^2$ and width $L/5$ in the center.  The new energy eigenstates and energy eigenvalues were determined numerically for this case.  We again consider an equal superposition of the ground state and first excited state, which is again periodic with angular frequency $(E_2'-E_1')/\hbar$.  Both energies are significantly less than $U_0$, so any flow across the barrier must be due to tunneling.  The situation is illustrated in Fig. \ref{fig:InfSqStreamsqr}, with streamlines indicating the motion of the fluid and shading indicating areas of superoscillation.

Considering the upper diagram in Fig. \ref{fig:InfSqStreamsqr}, we can see that the situation resembles the previous case without the barrier in that the cyclic motion is driven by a dip where the quantum potential becomes negative, which bounces back and forth between the two events where the nodes form.  The dip ($Q<0$, so soft superoscillation) creates a local boost in the flow kinetic energy $K_a$ as the fluid is pulled across, resulting in some regions of hard superoscillation, including the classically forbidden region within the barrier.  The quantum potential cancels out the external potential barrier, so the fluid flows through in almost the same way as if the barrier were absent, except that it is slower.

Considering the lower diagram in Fig. \ref{fig:InfSqStreamsqr}, in the $(Q_r, K_c)$ picture, we see the same role for the reduced quantum potential in driving the flow, but the wells are significantly deeper in some places, and some of the energy is stored locally in the form of symmetric kinetic energy $K_s$, so the regions of soft superoscillation where $Q_r<0$, or hard superoscillation where $K_c = K_a + K_s > E_2'$ are larger than in the standard $(Q,K_a)$ picture.

\subsection{Energy Eigenstates of the Harmonic Oscillator}
\begin{figure}
\centering
\begin{tabular}{c}
     \includegraphics[width=4.7in]{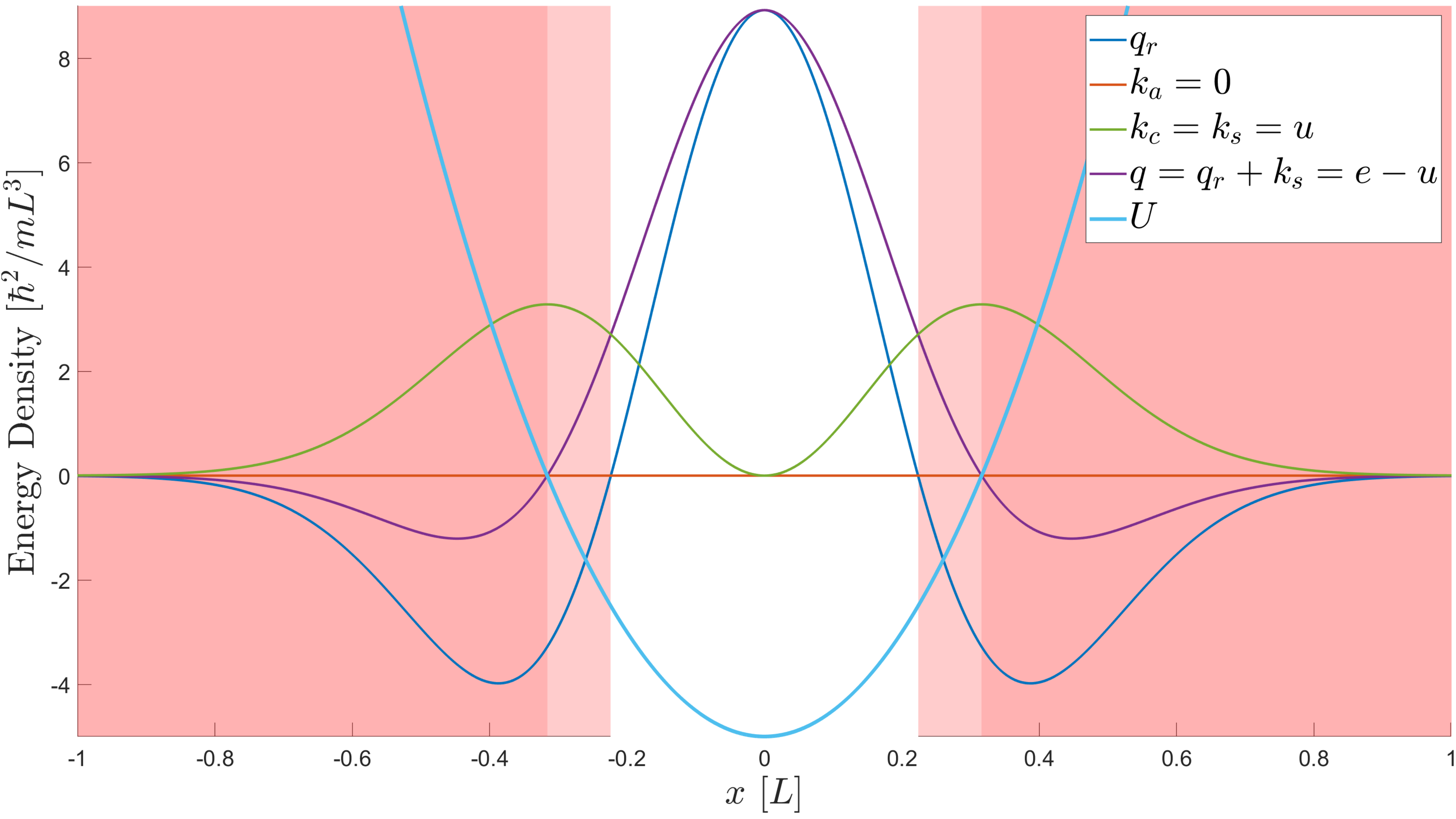}\\\\ \includegraphics[width=4.7in]{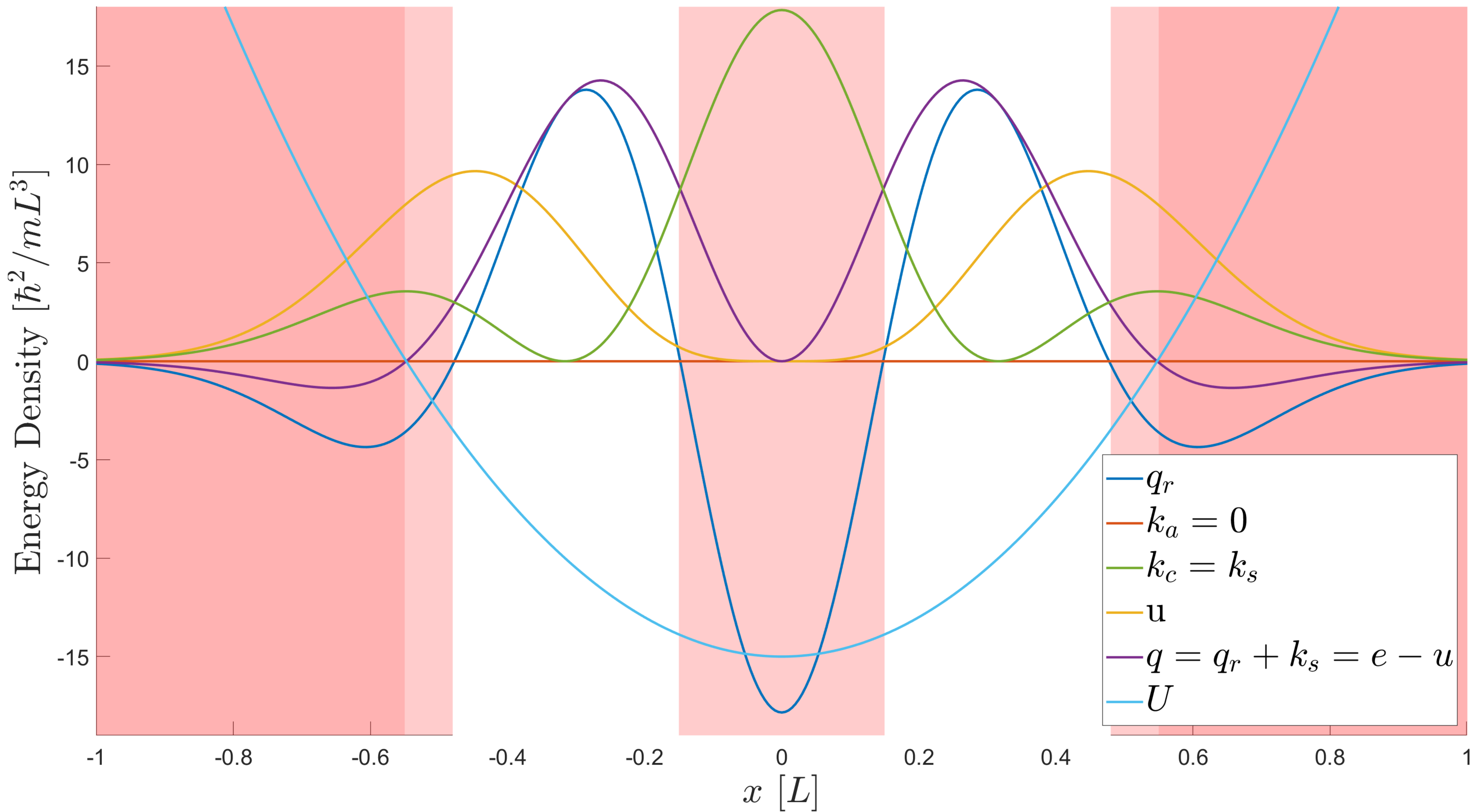}
\end{tabular}
    \caption{Energy densities $q$, $k_a$, $q_r$, $k_c$, $k_s$, and $v$ for the ground state (upper) and first excited state (lower) of the 1D harmonic oscillator.  The external potential $U$ is superimposed on the plot in arbitrary units, and the zero of the energy densities is set relative to $U$ at the energy eigenvalue.  The darker red shading shows regions where $q<0$ (hard/soft superoscillation in the ($q,k_a$) picture), which coincides with the classically forbidden region, and the lighter red shows regions where $q_r<0$ (in the ($q_r,k_c$) picture).  For the ground state $k_c = k_s = u$, so these are all plotted as one curve, whereas for the first excited state $k_c = k_s \neq u$.  Also, because $k_a=0$ for energy eigenstates, superoscillation in the $(q,k_a)$ picture occurs where $0 = k_a > e - u = q$, or simply $q<0$, where the central inequality shows kinetic energy exceeding the expected classical bound.  Likewise, for the ($q_r,k_c$) picture, superoscillation occurs where $k_s = k_c > e - u = q_r+k_s$, or simply $q_r<0$.}
    \label{fig:HOFG}
\end{figure}

The energy eigenstates of the 1D harmonic oscillator have fluid in the classically forbidden regions at the outside edges, which are associated with penetration depth rather than tunneling.  For eigenstates there is no difference between hard and soft superoscillations, so we depict both the ($Q,K_a$) and ($Q_r, K_c$) pictures in the same plots in Fig. \ref{fig:HOFG}, using different shades of red.  For these example plots, we used natural frequency $\omega = 10\hbar/mL^2$.  This example gives a clear idea of how ubiquitous superoscillations are in the physical world, since eigenstates of typical physical systems have tails in the classically forbidden region, which always corresponds to superoscillation.  In the $(Q_r, K_c)$ picture, there is superoscillation near all minima of the fluid density $R^2$ for all energy eigenstates, which also includes the classically forbidden regions.  Regions of superoscillation can be seen in the plots where the kinetic energy exceeds the expected classical kinetic energy, $k_a>e-u$ (also $q<0$), or $k_c>e-u$ (also $q_r<0$), in the respective pictures.

\subsection{Energy Superposition in the Harmonic Oscillator}

\begin{figure}
\centering
\begin{tabular}{c}
     \includegraphics[width=3.4in]{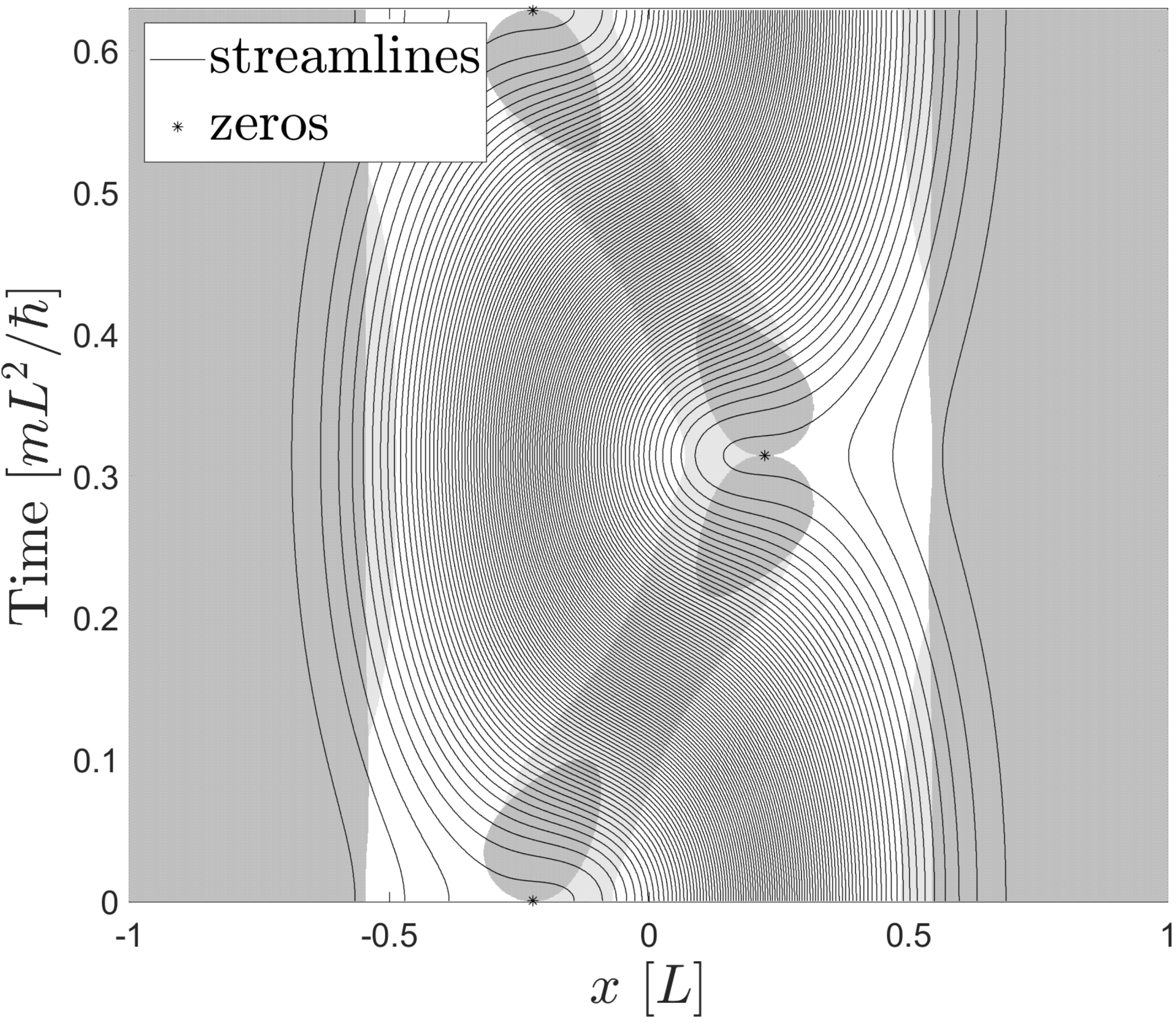}\\\\ \includegraphics[width=3.4in]{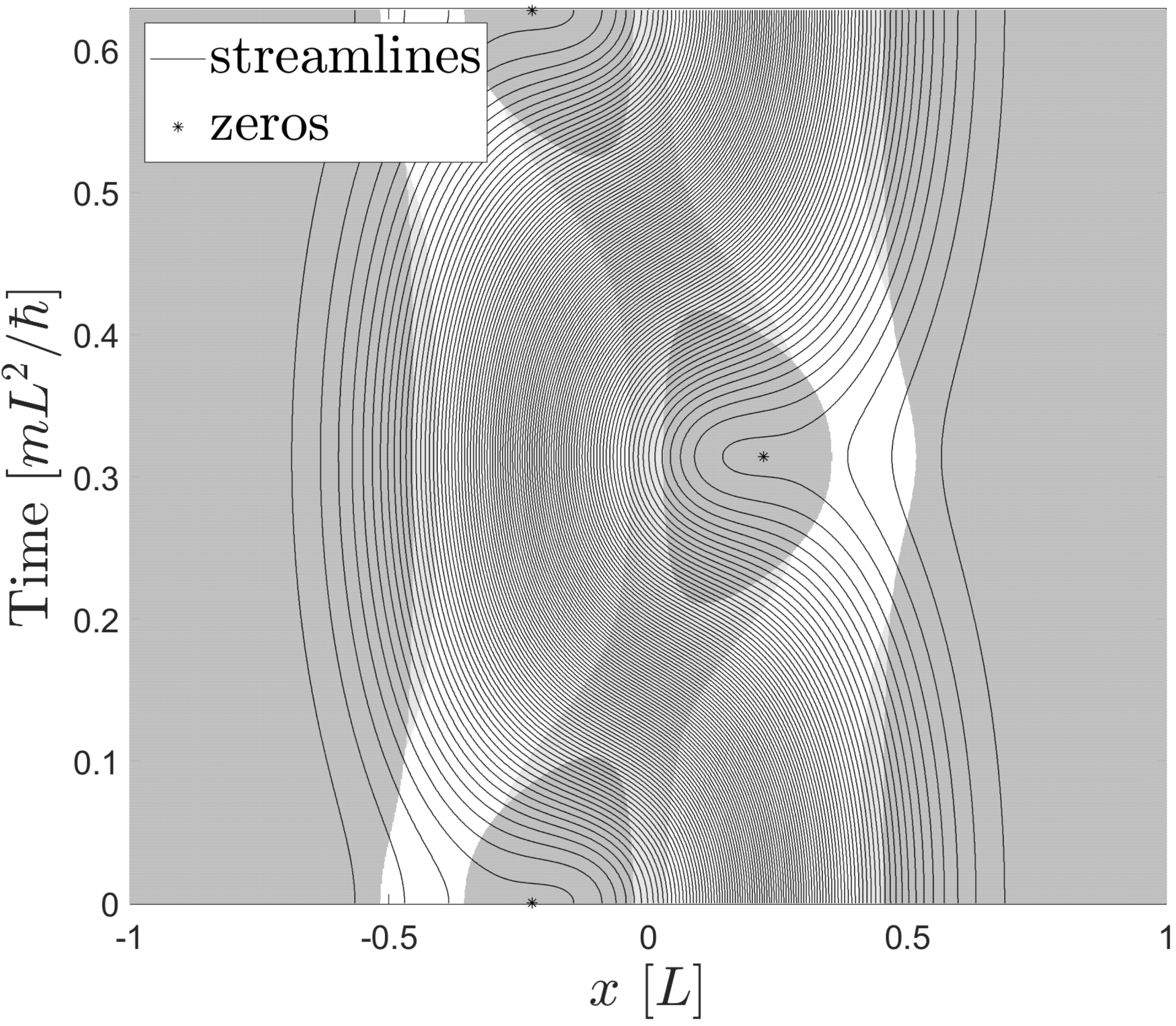}
\end{tabular}
    \caption{Streamline diagrams for one period of evolution for an equal superposition of the ground state and first excited state of a 1D harmonic oscillator with natural frequency $\omega = 10 \hbar/mL^2$.  Neighboring streamlines are separated by equal proportions of fluid, and so they depict the fluid density.  The nodes that appear once per cycle are shown with asterisks.  \textbf{upper:} The lighter gray shading indicates regions of soft superoscillation, $Q<0$ and the darker gray shading indicates regions of hard superoscillation $K_a > E_1-U$.  \textbf{lower:} The reduced quantum potential picture.  The darker gray shading indicates regions of soft superoscillation, $Q_r<0$ and the darker gray shading indicates regions of hard superoscillation $K_c >  E_1-U$.}
    \label{fig:HOStreams}
\end{figure}

We now consider an equal superposition of the ground state and first excited state of the 1D harmonic oscillator, which is periodic with angular frequency $(E_1-E_0)/\hbar$, with $E_0 = \hbar\omega/2$ and $E_1 = 3\hbar\omega/2$.  The situation is illustrated in Fig. \ref{fig:HOStreams}, with streamlines indicating the motion of the fluid and shading indicating areas of superoscillation.

Considering the upper diagram in Fig. \ref{fig:HOStreams}, we can see that the situation resembles the cases we saw for an energy superposition in the 1D infinite square well, with or without a finite barrier in the center, in that the cyclic motion is driven by a dip, where the quantum potential becomes negative, which bounces back and forth between the two events where the nodes form.  The dip ($Q<0$, so soft superoscillation) creates a local boost in the flow kinetic energy $K_a$ as the fluid is pulled across, resulting in some regions of hard superoscillation.  There is also the ubiquitous superoscillation in (or near) the classically forbidden region.

Considering the lower diagram in Fig. \ref{fig:HOStreams}, in the $(Q_r, K_c)$ picture, we see the same role for the reduced quantum potential in driving the flow, but the wells are significantly deeper in some places, and some of the energy is stored locally in the form of symmetric kinetic energy $K_s$, so the regions of soft superoscillation where $Q_r<0$, or hard superoscillation where $K_c = K_a + K_s > E_1$ are larger than in the standard $(Q,K_a)$ picture.

\subsection{Energy Eigenstates of a Quartic Double Well}

\begin{figure}
\centering
\begin{tabular}{c}
     \includegraphics[width=4.7in]{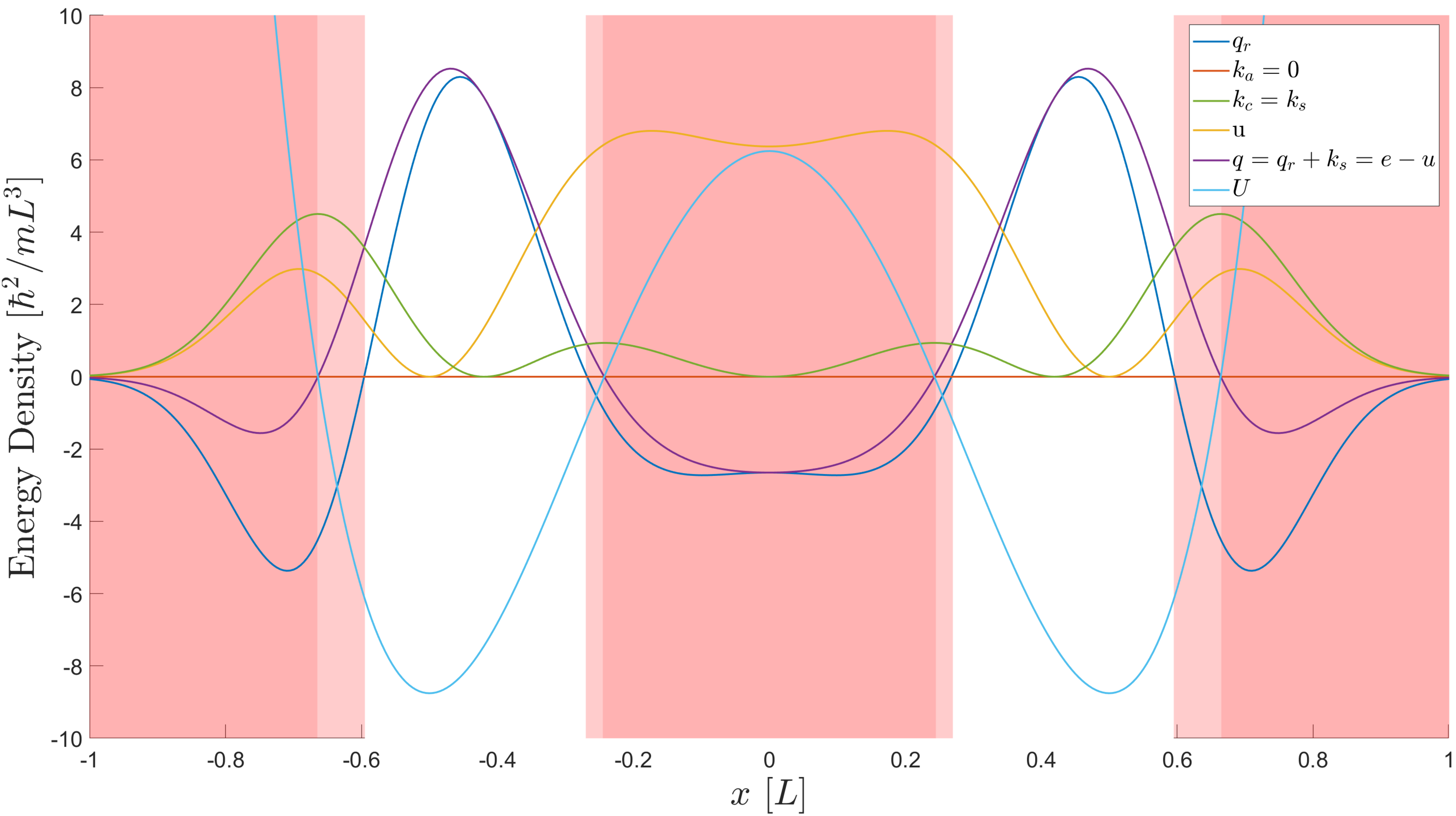}\\\\ \includegraphics[width=4.7in]{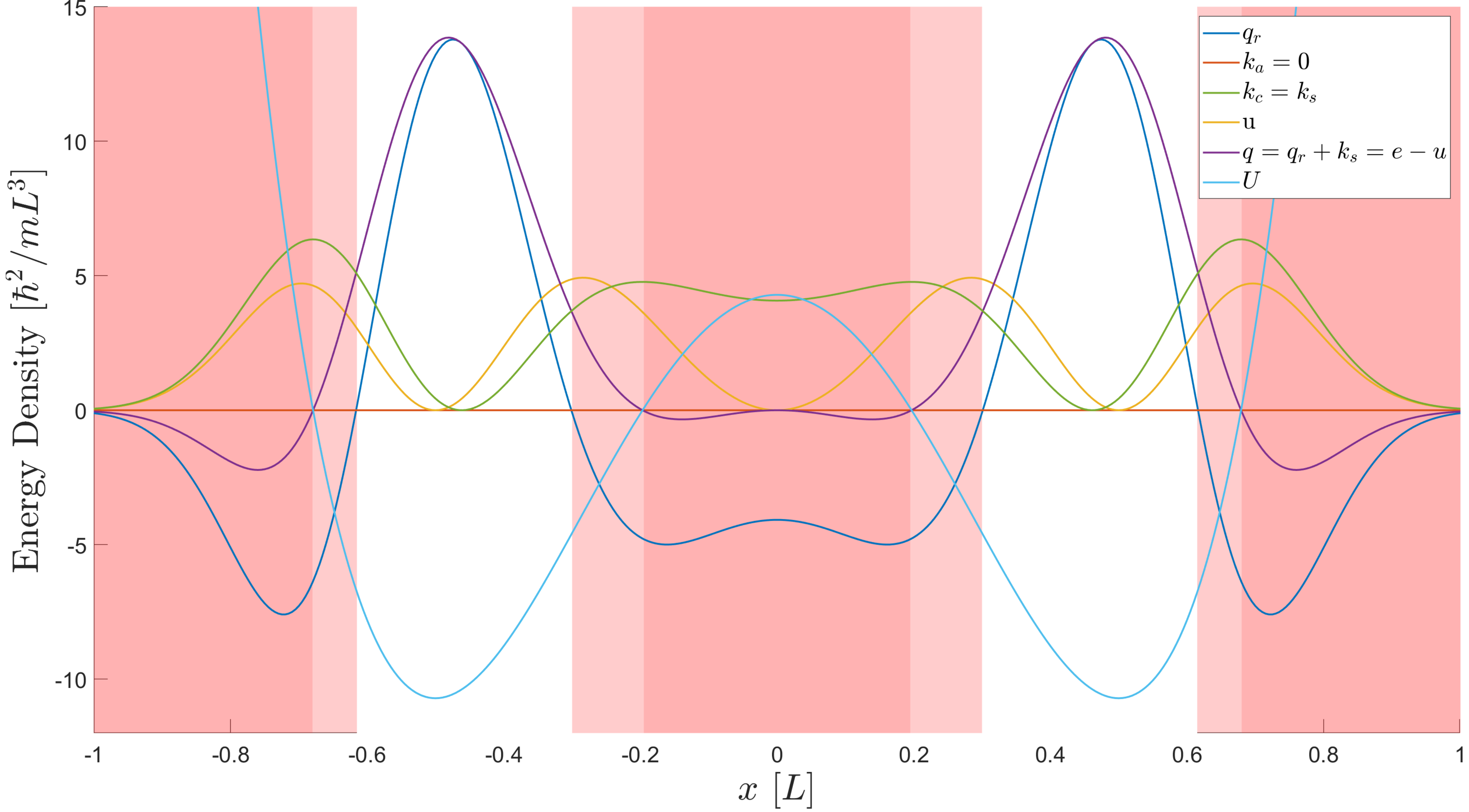}
\end{tabular}
    \caption{Energy densities $q$, $k_a$, $q_r$, $k_c$, $k_s$, and $v$ for the ground state (upper) and first excited state (lower) of a symmetric 1D quartic double well potential.  The external potential $U$ is superimposed on the plot in arbitrary units, and the zero of the energy densities is set relative to $U$ at the energy eigenvalue.  The darker red shading shows regions where $q<0$ (hard/soft superoscillation in the ($q,k_a$) picture), which coincides with the classically forbidden regions, and the lighter red shows regions where $q_r<0$ (in the ($q_r,k_c$) picture).  For the ground state $k_c = k_s = u$, so these are all plotted as one curve, whereas for the first excited state $k_c = k_s \neq u$.  Also, because $k_a=0$ for energy eigenstates, superoscillation in the $(q,k_a)$ picture occurs where $0 = k_a > e - u = q$, or simply $q<0$, where the central inequality shows kinetic energy exceeding the expected classical bound.  Likewise, for the ($q_r,k_c$) picture, superoscillation occurs where $k_s = k_c > e - u = q_r+k_s$, or simply $q_r<0$.}
    \label{fig:QDWFG}
\end{figure}

The energy eigenstates of the 1D quartic double well potential typically have fluid in the classically forbidden regions, both at the outside edges and within the central barrier.  For eigenstates there is no difference between hard and soft superoscillations, so we depict both the ($Q,K_a$) and ($Q_r, K_c$) pictures in the same plots in Fig. \ref{fig:QDWFG}, using different shades of red.  For these example plots, we used external potential energy,
\begin{equation}
    U(x) = \frac{\hbar^2}{mL^2}\bigg(240\frac{x^4}{L^4} - 120\frac{x^2}{L^2} + 15\bigg),  \label{QDW}
\end{equation}
which has minima $U(\pm 1/2) = 0$, and the barrier has peak height $U_0 = 15 \hbar^2/mL^2$.  The energy eigenstates and eigenvalues were found numerically.

This example is akin to the case we saw for the infinite square well with a finite barrier, but in a more physically realistic potential.    Again, we have superoscillation $Q<0$ in the classically forbidden regions.  In the $(Q_r, K_c)$ picture, there is superoscillation near all minima of the fluid density $R^2$ for all energy eigenstates, which also includes the classically forbidden regions.  Regions of superoscillation can be seen in the plots where the kinetic energy exceeds the expected classical kinetic energy, $k_a>e-u$ (also $q<0$), or $k_c>e-u$ (also $q_r<0$), in the respective pictures.

\subsection{Energy Superpositions in the Quartic Double Well}

\begin{figure}
\centering
\begin{tabular}{c}
     \includegraphics[width=5in]{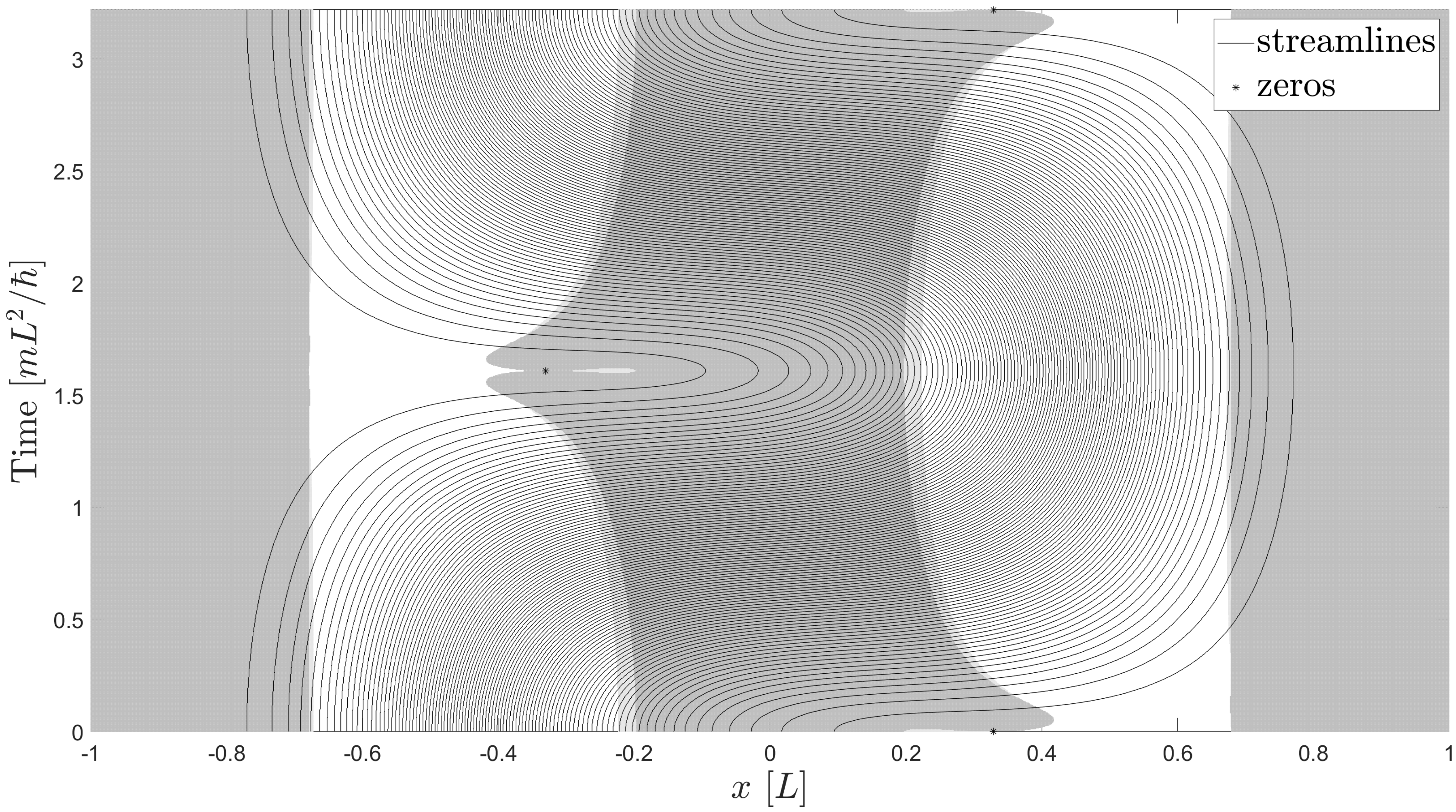}\\\\ \includegraphics[width=5in]{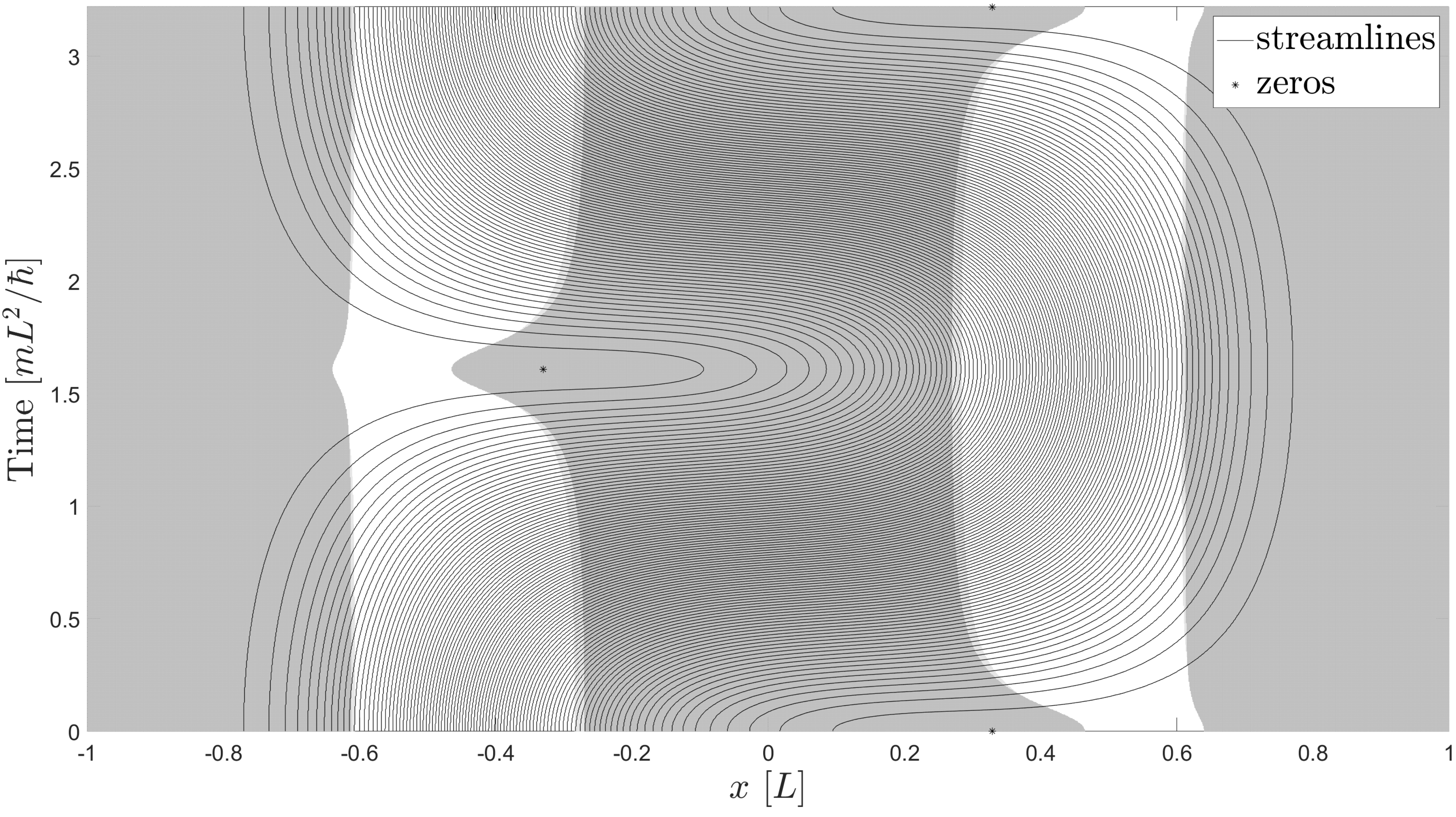}
\end{tabular}
    \caption{Streamline diagrams for one period of evolution for an equal superposition of the ground state and first excited state of a 1D quartic double well of Eq. \ref{QDW}.   Neighboring streamlines are separated by equal proportions of fluid, and so they depict the fluid density.  The nodes that appear once per cycle are shown with asterisks.  \textbf{upper:} The lighter gray shading indicates regions of soft superoscillation, $Q<0$ and the darker gray shading indicates regions of hard superoscillation $K_a > E_1-U$.  \textbf{lower:} The reduced quantum potential picture.  The darker gray shading indicates regions of soft superoscillation, $Q_r<0$ and the darker gray shading indicates regions of hard superoscillation $K_c >  E_1-U$.  The fluid tunnels through the barrier within continuous central region of superoscillation.}
    \label{fig:QDWStreams}
\end{figure}

We now consider an equal superposition of the ground state and first excited state of the 1D quartic double well potential $U(x)$ from the last section, which is periodic with angular frequency $(E_1-E_0)/\hbar$, with $E_0$ and $E_1$ are the numerically-determined energy eigenvalues of the ground state and first excited state, respectively.  Both energies are significantly less than $U_0$, so any flow across the barrier must be due to tunneling.  The situation is illustrated in Fig. \ref{fig:QDWStreams}, with streamlines indicating the motion of the fluid and shading indicating areas of superoscillation.  A full animation of this time evolution can be found in the supplementary Materials. 

Considering the upper diagram in Fig. \ref{fig:QDWStreams}, we see again that the cyclic motion is driven by a dip where the quantum potential becomes negative, which bounces back and forth between the two events where the nodes form.  The dip ($Q<0$, so soft superoscillation) creates a local boost in the flow kinetic energy $K_a$ as the fluid is pulled across, resulting in some regions of hard superoscillation.  As in the case of the infinite square well with a rectangular barrier, the quantum potential cancels out the barrier, and the flow looks quite similar to the case of the harmonic oscillator with no barrier. There is also the ubiquitous superoscillation in (or near) the classically forbidden region.

Considering the lower diagram in Fig. \ref{fig:QDWStreams}, in the $(Q_r, K_c)$ picture, we see the same role for the reduced quantum potential in driving the flow, but the wells are significantly deeper in some places, and some of the energy is stored locally in the form of symmetric kinetic energy $K_s$, so the regions of soft superoscillation where $Q_r<0$, or hard superoscillation where $K_c = K_a + K_s > E_1$ are larger than in the standard $(Q,K_a)$ picture.

\subsection{The Universality of Fluid Motion}

As we have seen in the four prior examples, the quantum potential energy always smooths out the external potential energy, such that the total potential energy is completely flat (constant) for energy eigenstates, $Q(\vec{x}) + U(\vec{x})= E$ (and $K_a = 0$), because the quantum potential exactly cancels all bumps and barriers in the external potential.  For energy superposition states, the total energy of a fluid particle is no longer constant, and their motion is guided by one or more moving valleys in the total potential $Q+U$, separating relatively flat hilltops where the fluid accumulates.  

\begin{figure}
    \centering
    \includegraphics[width=5in]{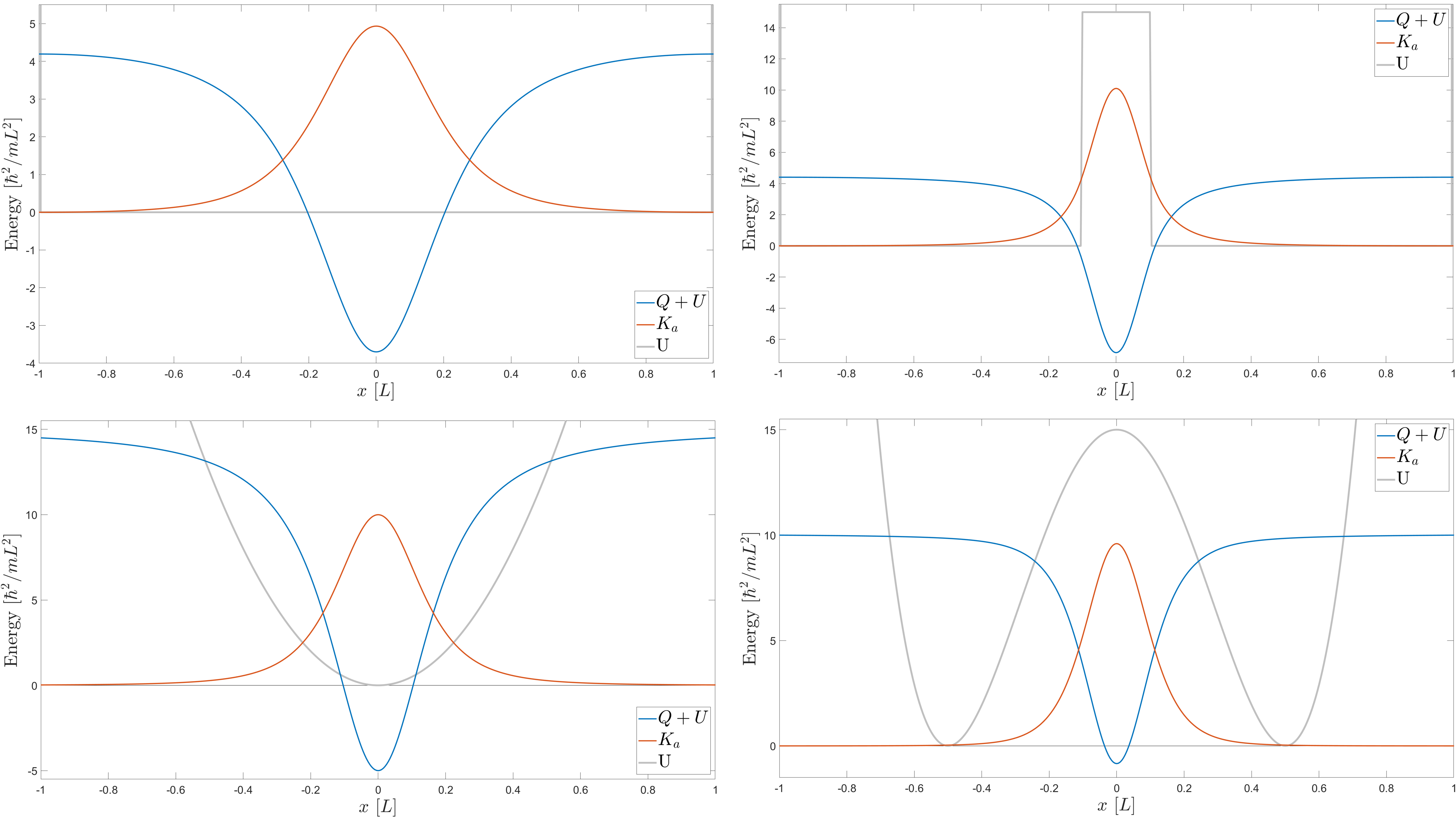}
    \caption{The kinetic energy $K_a$ and total potential energy $Q+U$ of an elemental particle in the fluid, for the four cases we have considered of an equal superposition of the ground state and first exited state, in four different binding potentials $U$ (\textbf{upper left:} infinite square well, \textbf{upper right:} infinite square well with a rectangular barrier, \textbf{lower left:} harmonic oscillator, \textbf{lower right:} quartic double well), at time $t = T/4$, where $T$ is the period of oscillation of the corresponding state (see the full animation in the Supplementary Materials).  At this moment, in all four cases, the fluid's density is symmetric about $x=0$, it is halfway through flowing from one side of the well to the other, and its net kinetic energy is maximum.  Also, the shape of $K_a$ and $Q+U$ looks essentially the same for all four cases, even though two of them are in the middle of tunneling through a barrier, and two of them are open wells.  This basic shape is retained as the dip travels between the nodes in the fluid density, but changes as it approaches or departs a node.  The dip grows deeper as it approaches the locations where a node forms, and a positive bump begins to appear on its leading edge.  For $Q$, the dip extends instantaneously to $-\infty$ where the singularity forms, and the bump on the leading edge extends instantaneously to $+\infty$, while for $K_a$, the spike extends instantaneously to $+\infty$.  The fluid density is zero at these events, so no elemental fluid particles have these infinite energies, but the particles very close to the node do have arbitrarily large finite energies.}
    \label{fig:QKA4}
\end{figure}

Remarkably, this picture appears to be universal.  Regardless of the external potential, its shape is canceled by the quantum potential, and the moving hills and valleys in the quantum potential guide all of the kinetic motion of the fluid (for static external potentials).  To see this, consider Fig. \ref{fig:QKA4}, which shows the flow kinetic energy $K_a$ and the total potential energy $Q+U$ of an elemental particle in the fluid, for the four cases we have considered of an equal superposition of the ground state and first exited state, in four different binding potentials $U$, at time $t =T/4$, where $T$ is the period of oscillation for the given state.

\subsection{Interference During Reflection}

\begin{figure}
\centering
\begin{tabular}{c}
     \includegraphics[width=4.5in]{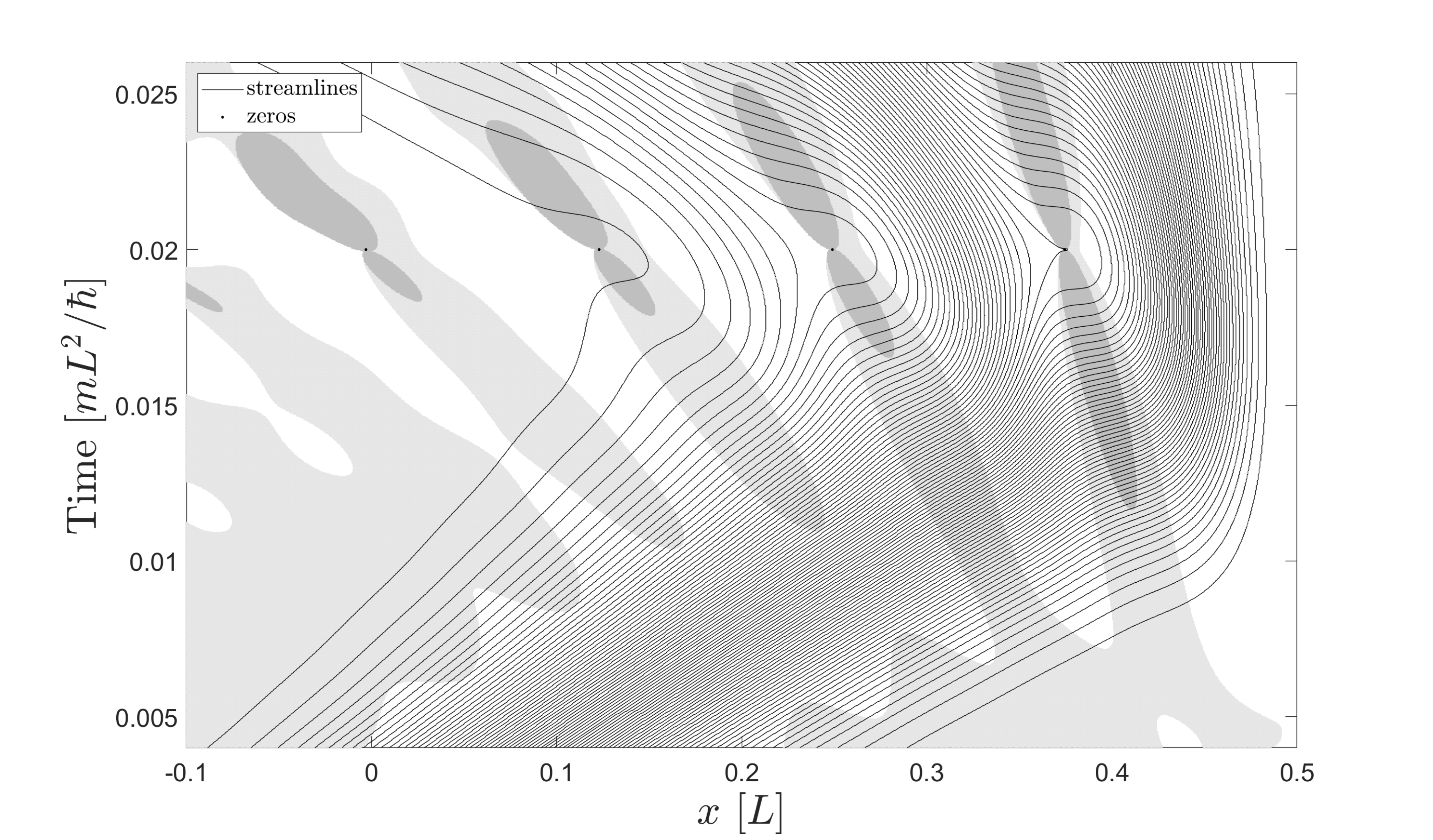}\\\\ \includegraphics[width=4.5in]{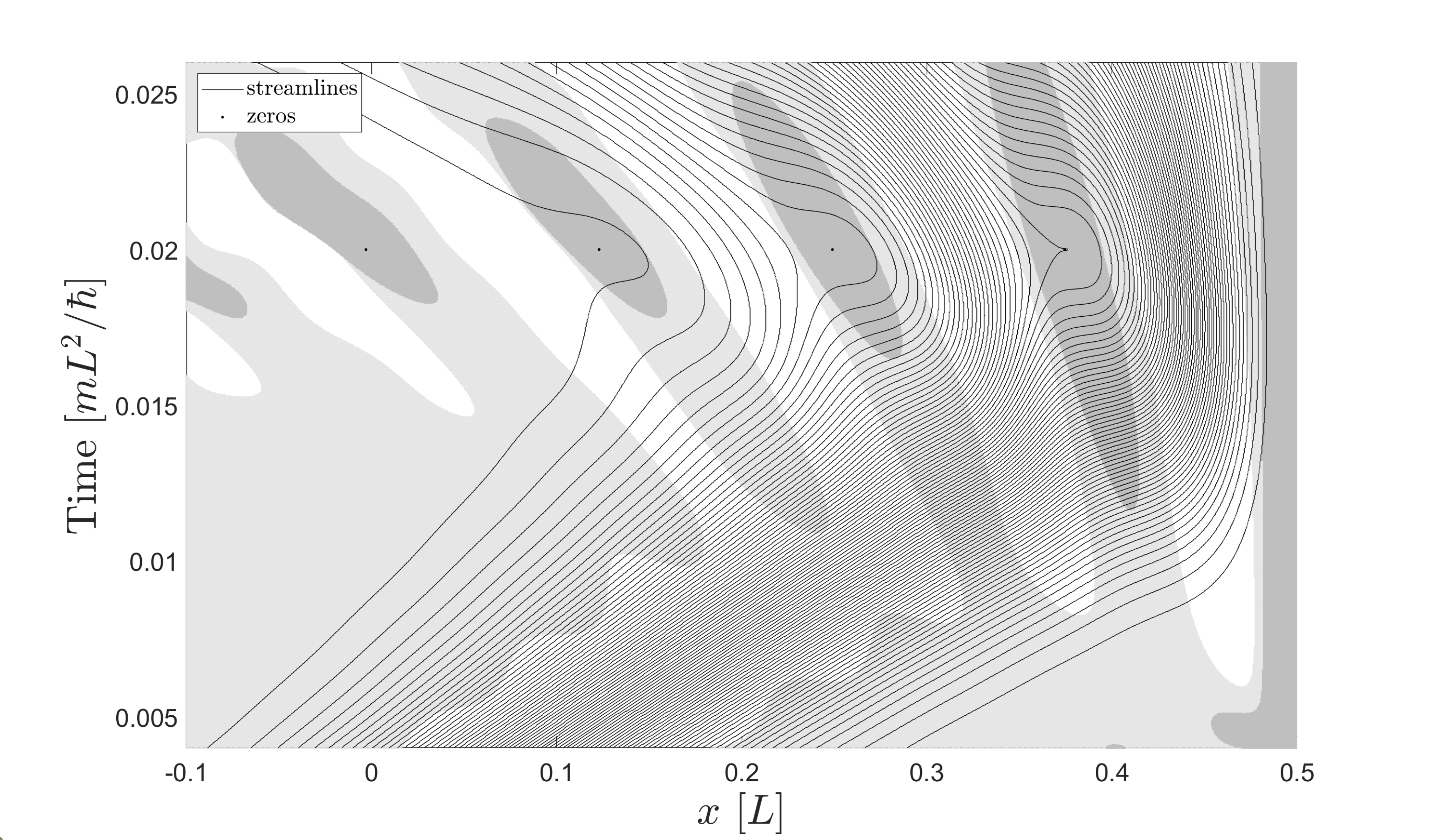}
\end{tabular}
    \caption{Streamline diagrams for a nearly-Gaussian band-limited wave pulse reflecting off an infinite wall and interfering with itself.  The pulse also spreads significantly during this process.  Neighboring streamlines are separated by equal proportions of fluid, and so they depict the fluid density.  The nodes that appear half-way through the reflection are shown with dots, and one fluid streamline can be seen whipping around the deep pit in $Q$ that forms there.  
    As in the simpler examples, we can see that the fluid motion is guided by a series of moving hills and valleys in the quantum potential which produce the separation into distinct interference fringes.  \textbf{upper:} The lighter gray shading indicates regions of soft superoscillation, $Q<0$ and the darker gray shading indicates regions of hard superoscillation $K_a > E_1-U$.  \textbf{lower:} The reduced quantum potential picture.  The darker gray shading indicates regions of soft superoscillation, $Q_r<0$ and the darker gray shading indicates regions of hard superoscillation $K_c >  E_1-U$.}
    \label{fig:Reflection}
\end{figure}

We now move on to consider a nearly-Gaussian band-limited pulse as it reflects off an infinite wall.  We begin at $t=0$ with a Gaussian of momentum $25\hbar/L$ centered at $x=0$ within an infinite square well.  This initial wavefunction is expanded into the eigenbasis $\psi_n(x)$ of an infinite square well of width $L$.  The coefficients $c_n$ in this expansion become negligible for $n>18$, so we obtain our band limited wave pulse by only using the first eighteen terms in the expansion, and renormalizing.  Fig. \ref{fig:Reflection} shows the fluid streamlines as this pulse reflects off the wall of the well and interferes with itself.  The nodes are labeled, and shading shows regions of hard and soft superoscillation.

\subsection{A 1D Mach-Zehnder Interferometer}

\begin{figure}
\centering
\begin{tabular}{c}
     \includegraphics[width=4.9in]{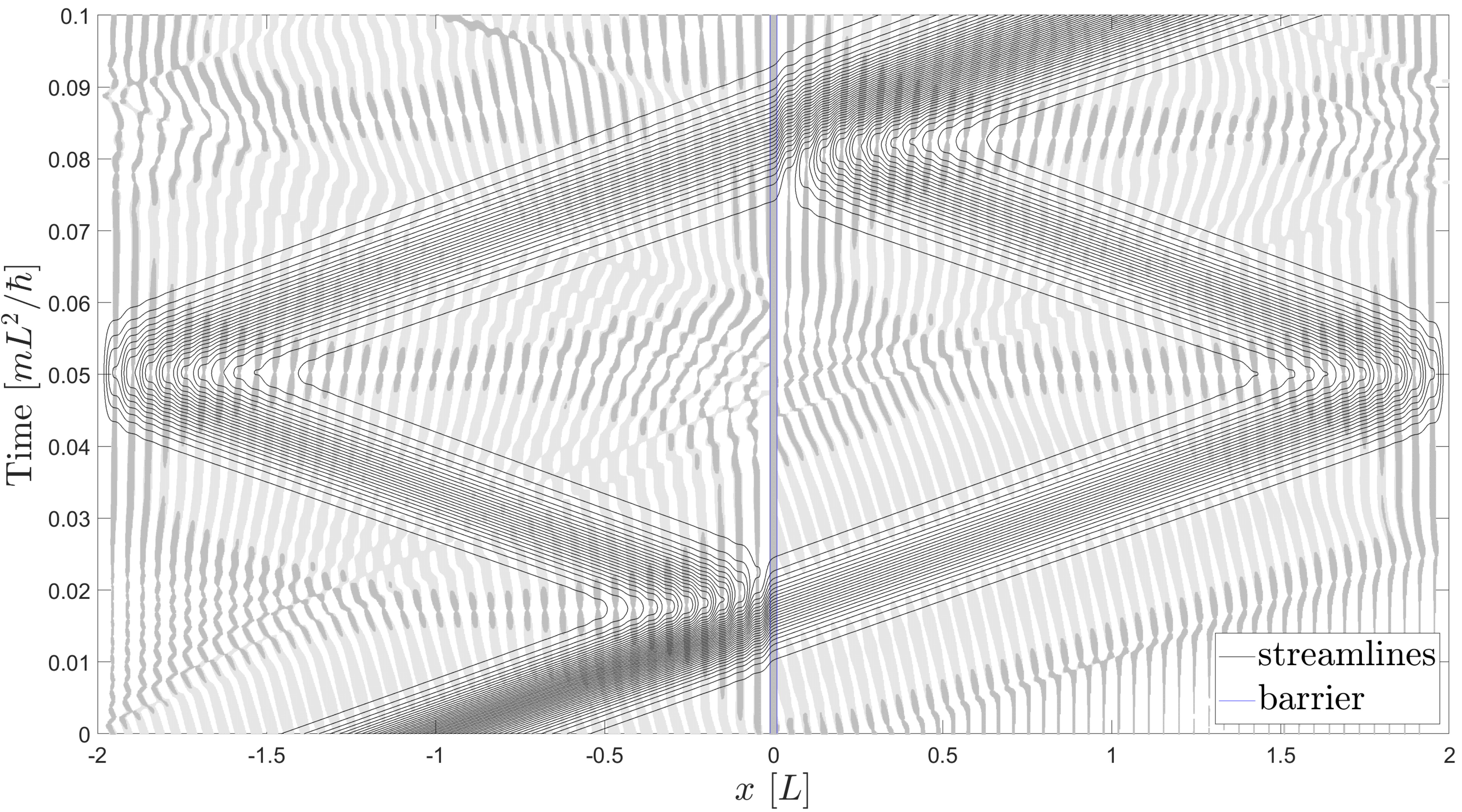}\\\\ \includegraphics[width=4.9in]{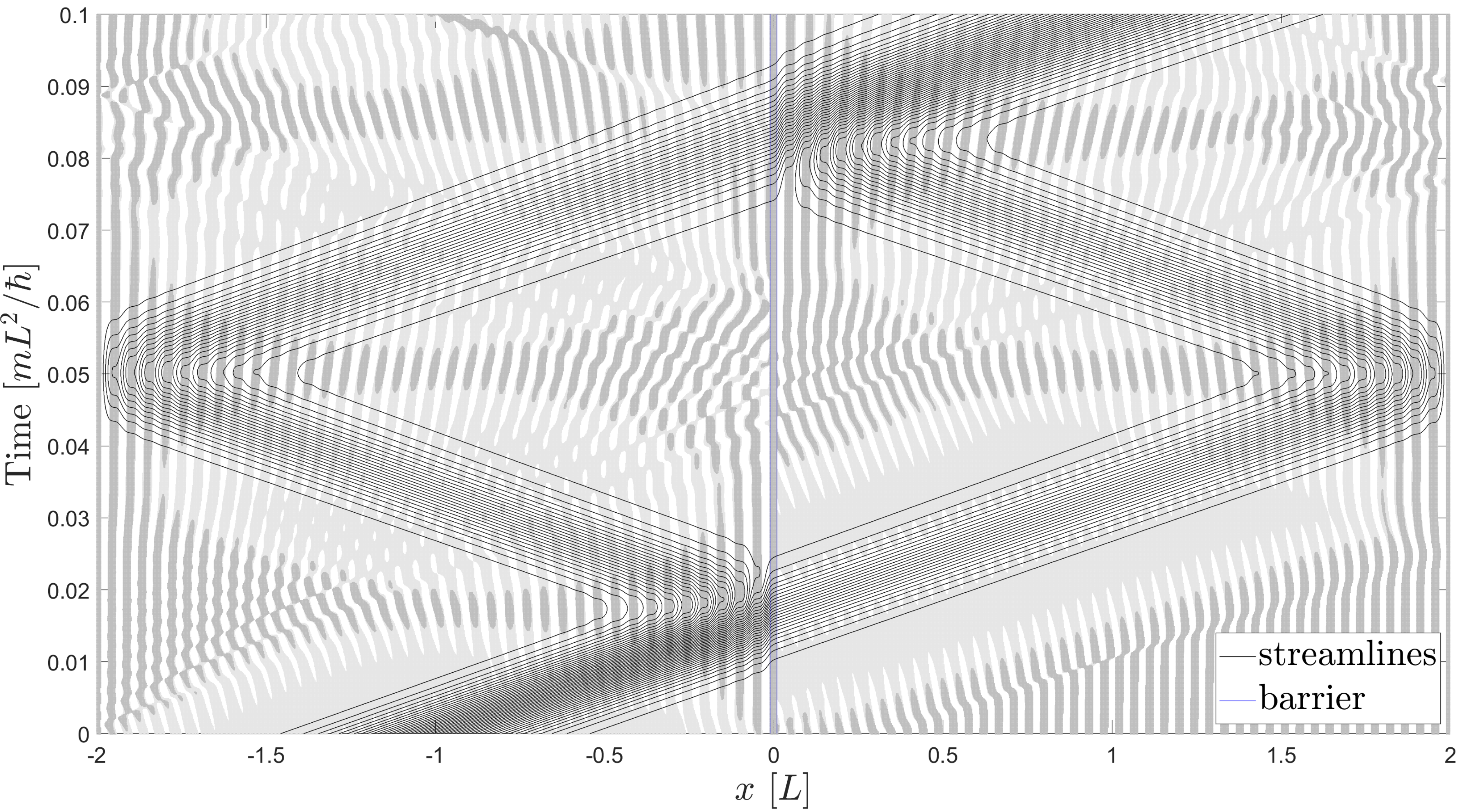}
\end{tabular}
    \caption{Streamline diagrams for a nearly-Gaussian band-limited wave pulse dividing at beam splitter, reflecting off an infinite wall and recombining at the beam splitter due to interference, effecting a 1D Mach Zehnder Interferometer.  Neighboring streamlines are separated by equal proportions of fluid, and so they depict the fluid density.  As in the simpler examples, we can see that the fluid motion is guided by a series of moving hills and valleys in the quantum potential which produce the separation into distinct interference fringes.  \textbf{upper:} The lighter gray shading indicates regions of soft superoscillation, $Q<0$ and the darker gray shading indicates regions of hard superoscillation $K_a > E_1-U$.  \textbf{lower:} The reduced quantum potential picture.  The darker gray shading indicates regions of soft superoscillation, $Q_r<0$ and the darker gray shading indicates regions of hard superoscillation $K_c >  E_1-U$.}
    \label{fig:MZ}
\end{figure}

So far, the examples we have considered have been simple cases at relatively low energy, which help to reveal the underlying mechanisms at work in the fluid interpretation, but it is also illustrative to consider a familiar higher-energy scenario, which is a 1D Mach-Zehnder interferometer (MZI).  This is an infinite square well of width $4L$ with a thin barrier of width $L/50$ and height $U_0$ in the center at $x=0$.  We again construct a Gaussian pulse with an initial positive momentum $60\hbar/L$, which begins centered at $x=-L$, and expand this into the energy eigenbasis of the infinite square well.  Then, noting that the coefficients $c_n$ become negligible for $n>89$, we get a band-limited pulse by using only the first 89 terms in the expansion, and renormalizing.  The barrier height $U_0$ was then tuned so that exactly half of the band-limited pulse is reflected and half is transmitted, resulting in a 50/50 beam splitter, although one which, due to the barrier width, slightly distorts the wave packet.  Fig. \ref{fig:MZ} shows the packet as it is divided, reflected, and then recombined (almost perfectly), and thus we effectively have a MZI in a 1D cavity, with shading showing the regions of soft/hard superoscillation.

Two zoomed-in animations of a band limited Gaussian pulse incident on different finite barriers tuned as a 50/50 beam splitter are given in the Supplementary Materials.  The parameters are chosen differently than in the 1D MZI shown here in order to better illustrate the details of tunneling and interference in this model.

\subsection{Degenerate Superposition in the 2D Infinite Square Well} \label{Vortex}

\begin{figure}
    \centering
    \includegraphics[width=5in]{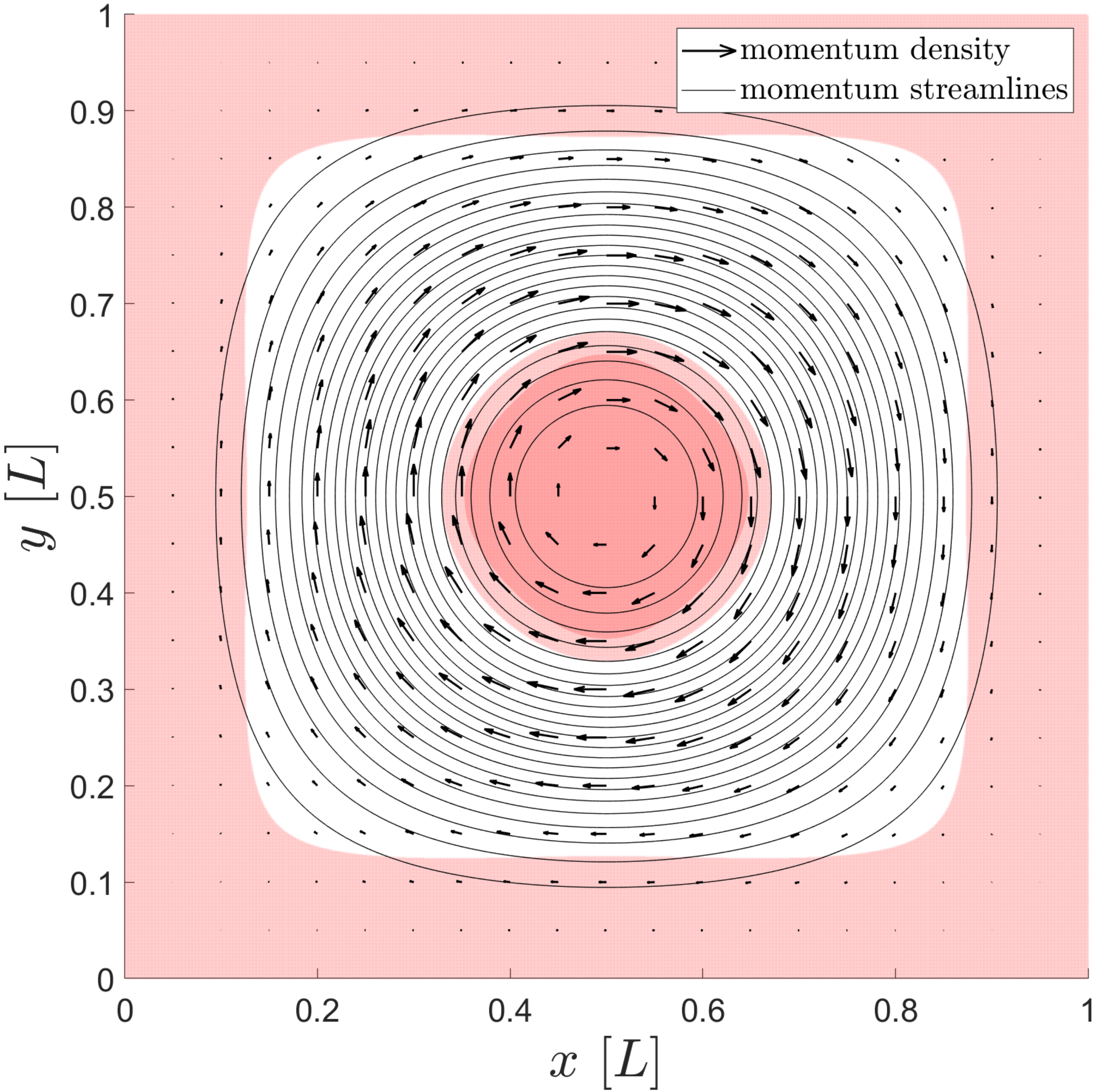}
    \caption{The momentum density and streamlines of the degenerate superposition state of Eq. \ref{psi2D} in the 2D infinite square well, with side lengths $L$. The momentum streamlines start separated by equal proportions of fluid along the vertical axis of symmetry, and then closed loops are generated by following the momentum flow from those starting points.  The arrows show the local direction and magnitude of the momentum density.  In the $(Q, K_a)$ picture, we have a stable vortex with the fluid orbiting with kinetic energy $K_a$ around a central singularity in $Q$.  The region shaded a darker red has $Q<0$, which shows superoscillation in the $(Q,K_a)$ picture, and the region shared the lighter red has $Q_r<0$ also shows superoscillation in the $(Q_r, K_c)$ picture.}
    \label{fig:2DMomentumStreams}
\end{figure}

\begin{figure}
    \centering
    \includegraphics[width=5in]{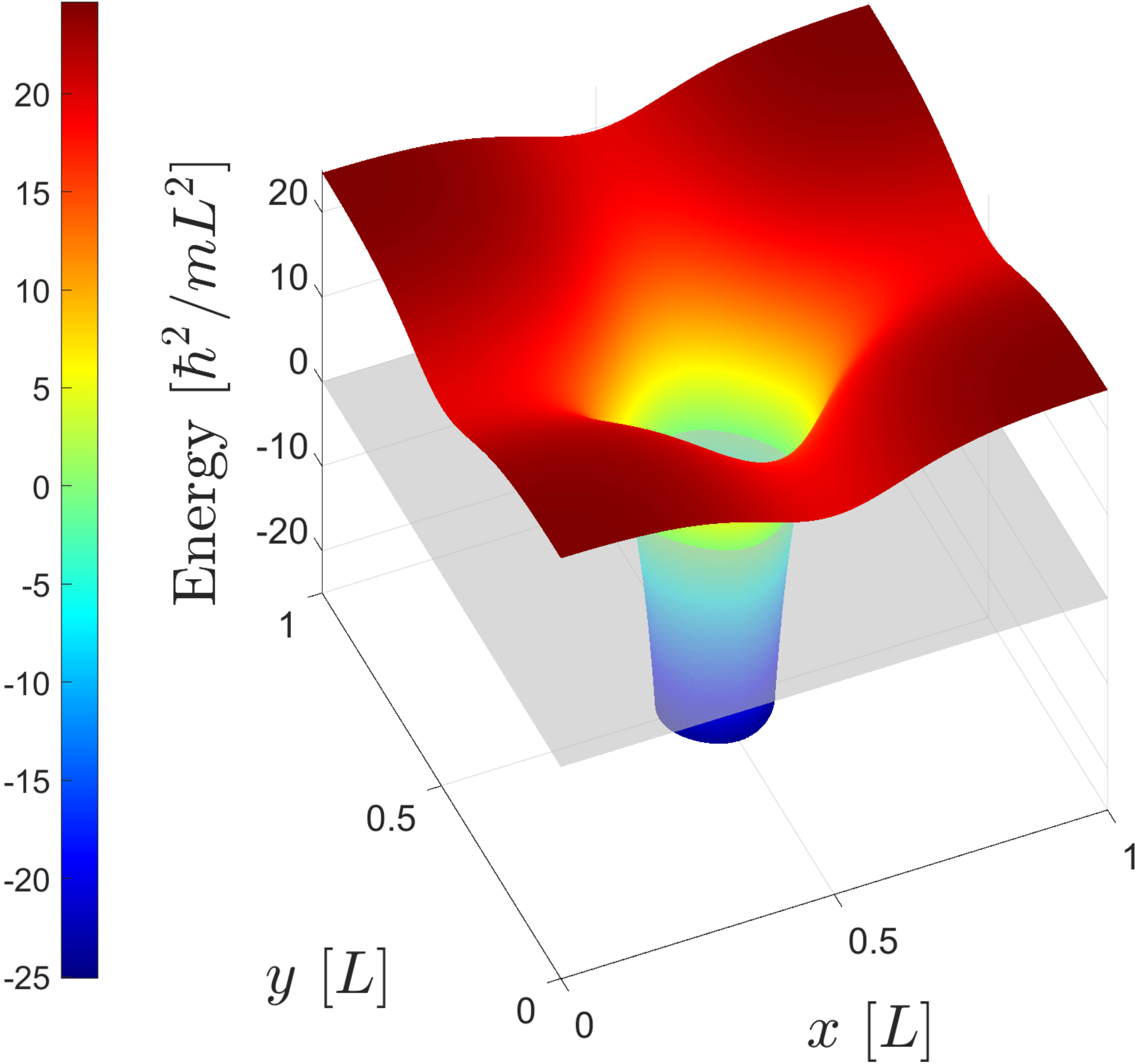}
    \caption{The quantum potential energy $Q$ in which the fluid particles orbit (the external potential is $U=0$ inside the well).  In the $(Q,K_a)$ picture, these are essentially classical orbits with speed $v_a = (\vec{\nabla}S)/m $ an a force $-\vec{\nabla}Q$, and near the center of the well, they become uniform circular orbits, and the quantum potential goes as $-1/r^2$, where $r$ is the distance from the center.  Regions where $Q$ is below the translucent gray plane at $Q=0$ are superoscillatory, where the orbital speeds within the vortex approach infinity.  In the $(q_r, k_c)$ picture, some of the $Q$ seen here is understood as symmetric kinetic energy $K_s$ which properly belongs to the total classical kinetic energy $K_c$, but it is still the total $Q = Q_r+K_s$ that guides the bulk flow with energy $K_a$.}
    \label{fig:2DQFunnel}
\end{figure}

\begin{figure}
 \centering
\begin{tabular}{c|c}
Quantum Potential, $q$ & Average Kinetic, $k_a$
\\
\includegraphics[width=2.35in]{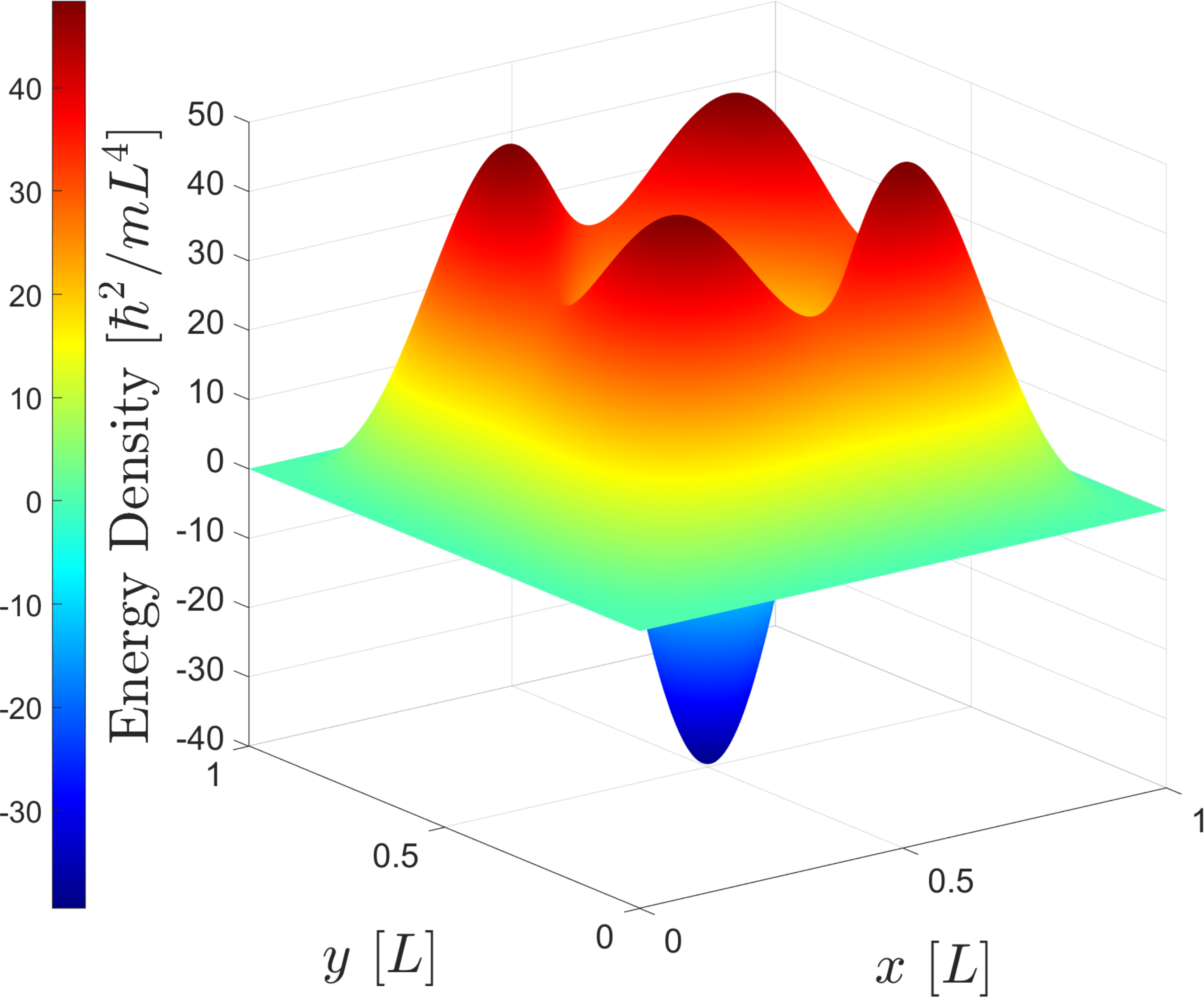} &  \includegraphics[width=2.35in]{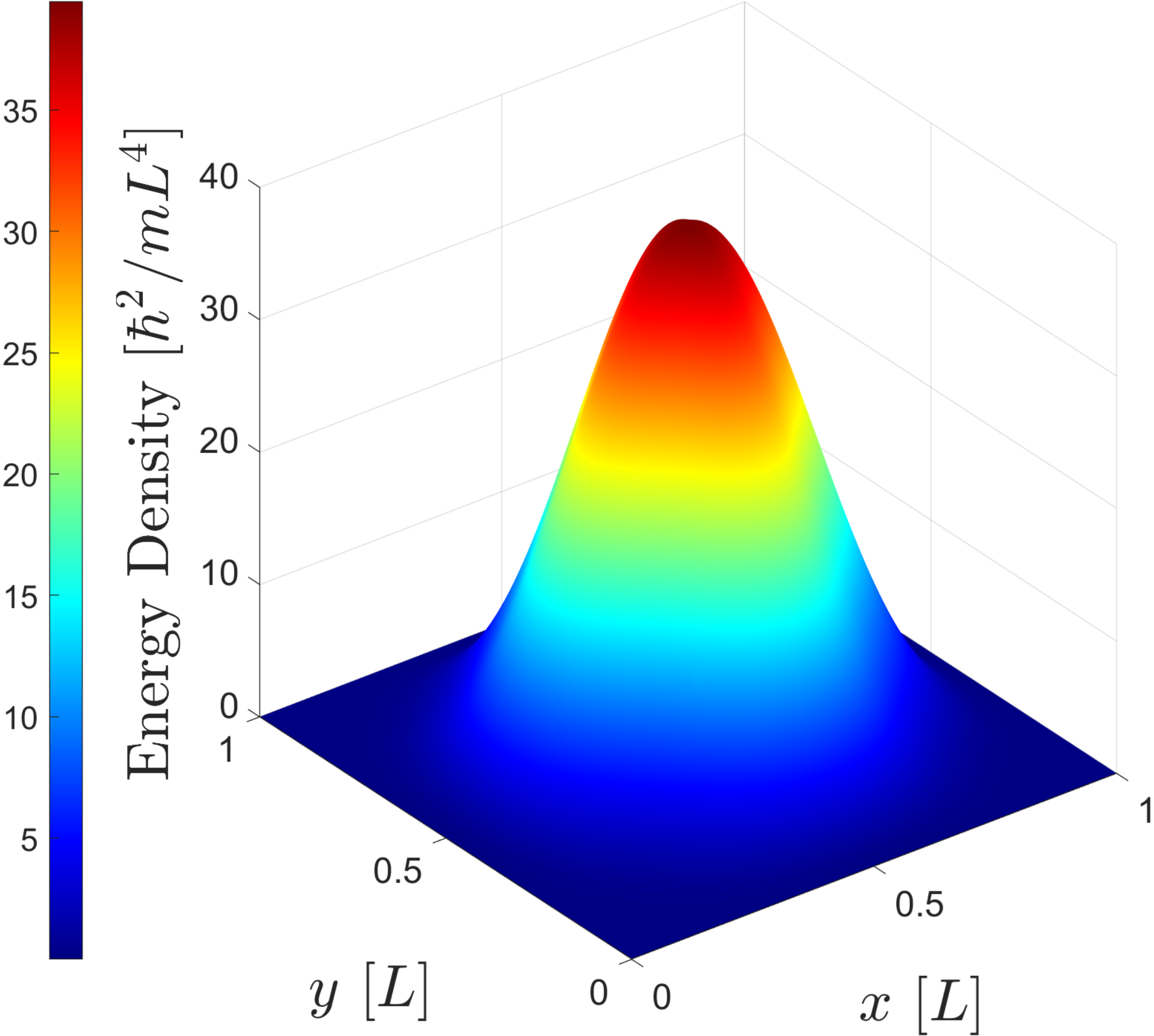}\\  
\hline\\
Reduced Quantum Potential, $q_r$ & Classical Kinetic, $k_c$ \\
\includegraphics[width=2.35in]{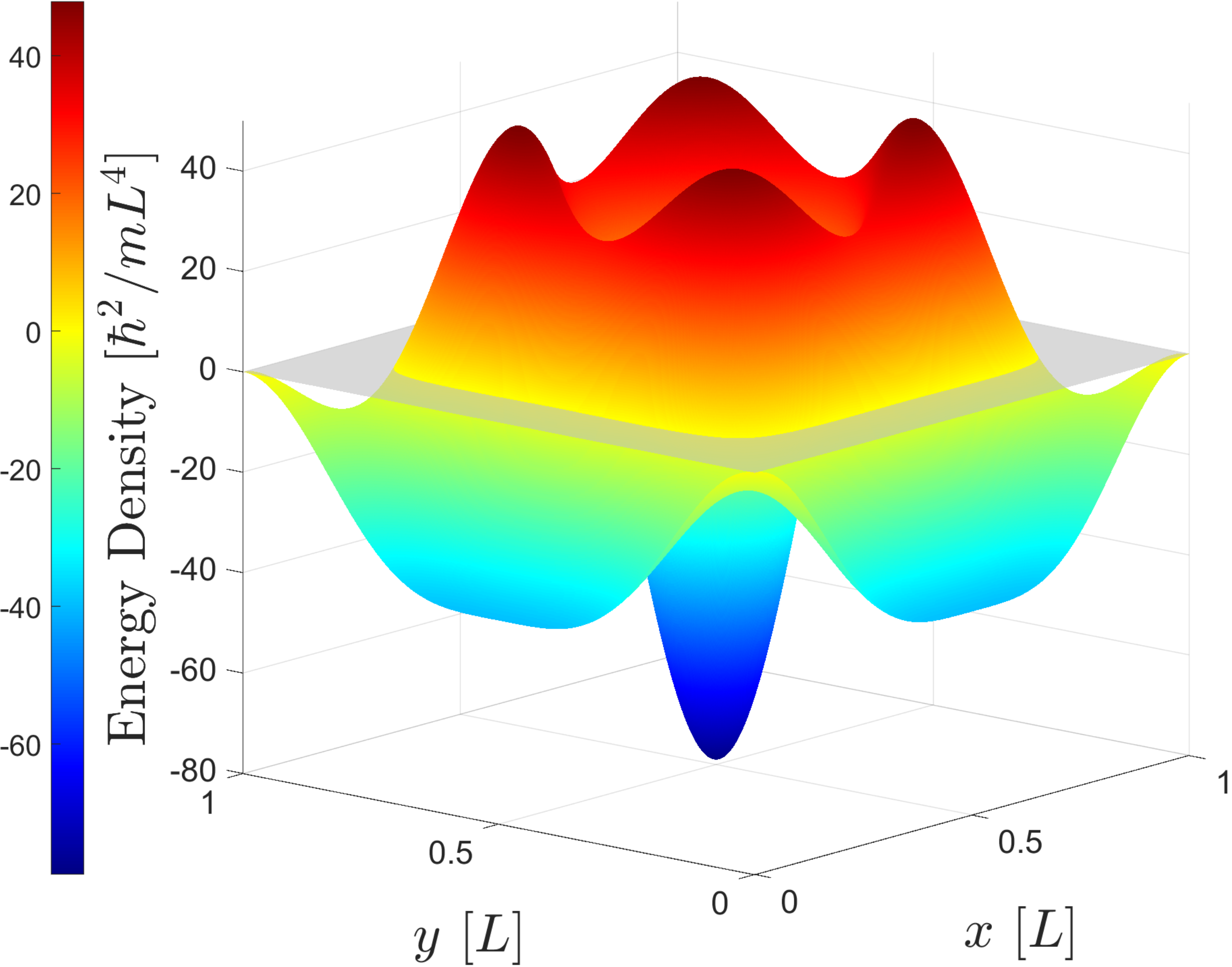} 
     & \includegraphics[width=2.35in]{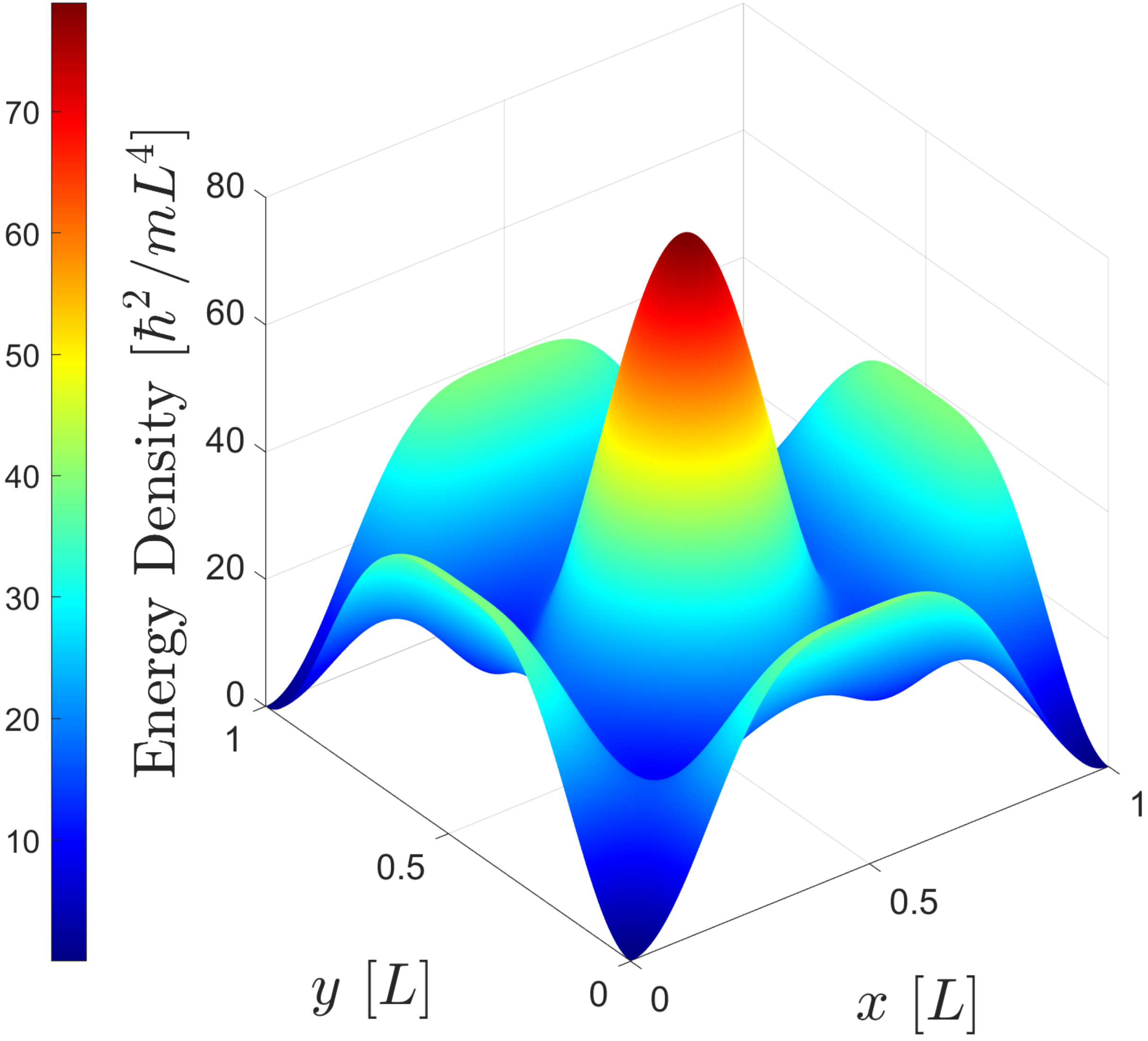} \\
\hline\\
Fluid Density, $\rho\equiv R^2 \equiv |\psi|^2$  &  Symmetric Kinetic, $k_s$ \\
     \includegraphics[width=2.35in]{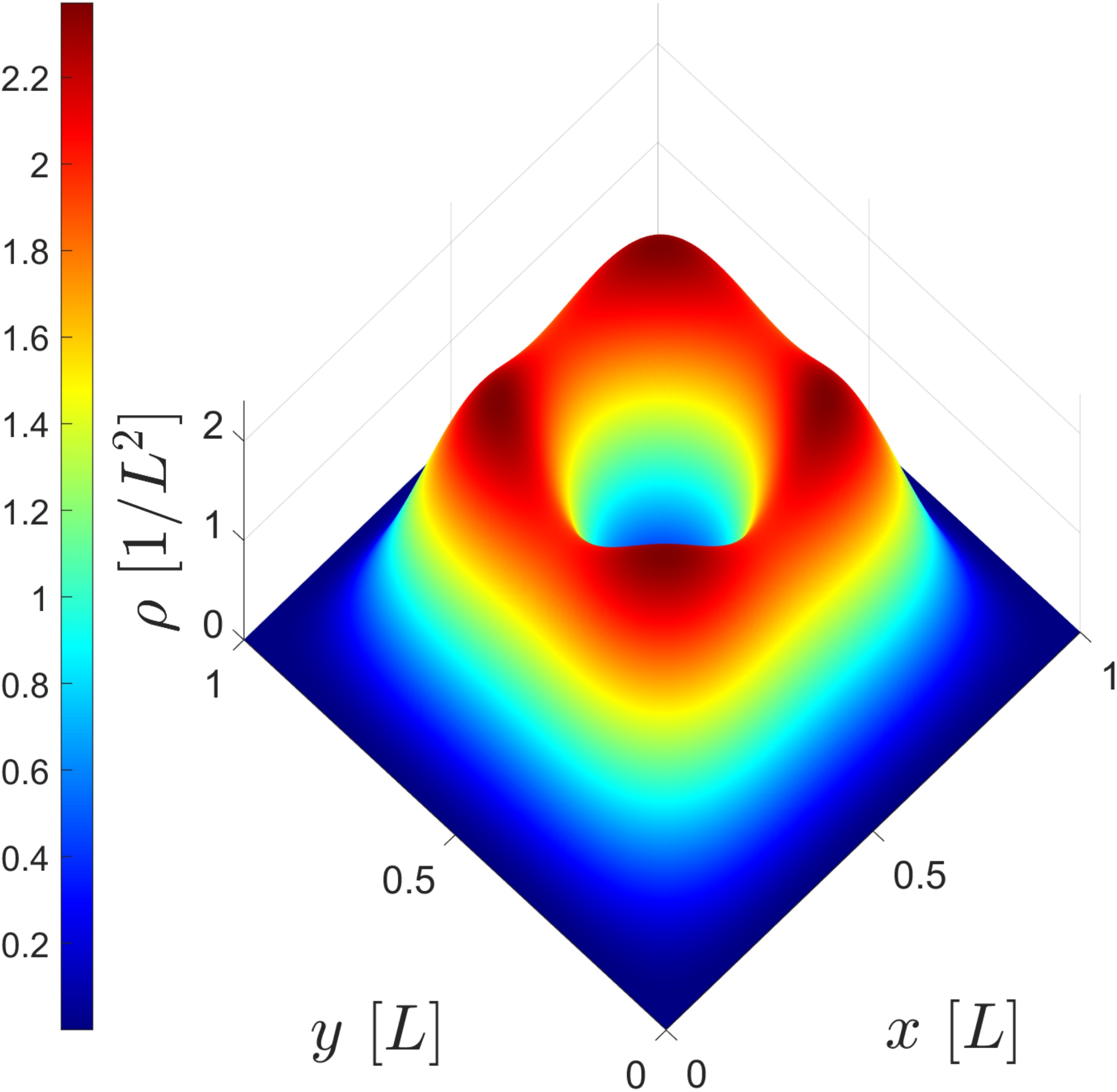} &  \includegraphics[width=2.35in]{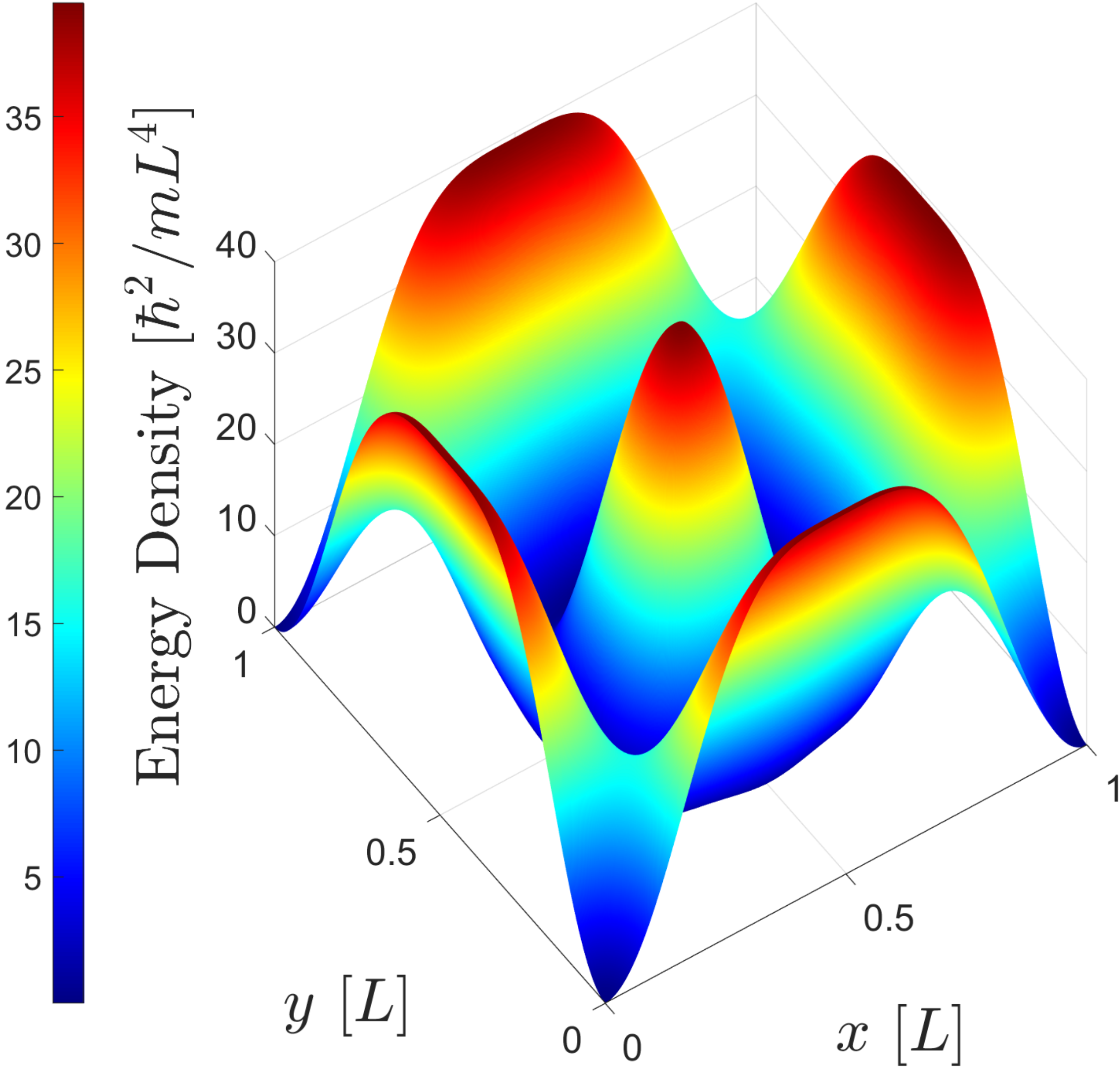}
\end{tabular}

    \caption{Fluid and energy densities for the state of Eq. \ref{psi2D} in the 2D infinite square well.}
    \label{fig:psi2Ddensities}
\end{figure}

Another important example can only be seen in 2D or 3D, and this is a complex superposition of degenerate eigenstates.  We consider the 2D infinite square well with side lengths $L$, and the degenerate eigenvalues are $(n_x=1,n_y=2)$ and $(n_x=2,n_y=1)$ superposition,
\begin{equation}
    \psi(x,y,t) = \frac{\sqrt{2}}{L}e^{-\frac{iEt}{\hbar}}\bigg(\sin\Big( \frac{\pi x}{L} \Big)\sin\Big( \frac{2\pi y}{L} \Big) + i\sin\Big( \frac{2 \pi x}{L} \Big)\sin\Big( \frac{\pi y}{L} \Big) \bigg), \label{psi2D}
\end{equation}
    which has energy $E = \frac{5\pi^2 \hbar^2}{ 2 m L^2}$.  Because this state has definite energy, there is no difference here between hard and soft superoscillation (i.e., $-\frac{\partial S}{\partial t} = E$).  However, because the superposition is complex, it is not the case that $\vec{\nabla}S=0$, which means the fluid is in motion, even though the fluid density $|\psi|^2 \equiv \rho \equiv R^2$ is constant, so we still have a stationary wave.  In this case, the fluid moves in stable orbits around the center of the well, where a node in the density results in a negative singularity in the quantum potential, maintaining a symmetric vortex distribution.  Near the singularity, the orbits are essentially circular, which means that the quantum potential energy $Q$ must be proportional to $-1/r^2$, with $r$ the distance from the center, as discussed above.  The momentum density and streamlines are shown in Fig. \ref{fig:2DMomentumStreams}, and regions of $(q,k_a)$-picture superoscillation, $Q<0$. are shown in darker shaded red, and regions of $(q_r, k_c)$-picture superoscillation, $Q_r<0$ are shown in lighter red.  The shape of the quantum potential energy in which the fluid particles orbit is shown in Fig. \ref{fig:2DQFunnel}. 

The fluid density $\rho$ of the stable vortex, and the five energy densities of interest are shown in Fig. \ref{fig:psi2Ddensities}.  It is worth noting that $k_a$ is not zero at the node in $\rho$, and is in fact at a local maximum there.  In our other examples, this was seen for $k_s$, but $k_a=0$ at the nodes.  The similarity between this case for $k_a$ and the more general case for $k_s$ further motivates the interpretation of $k_s$ as describing orbital motion about axis $\hat{v}_s(\vec{x},t)$ at speed $|\vec{v}_s(\vec{x},t)|$ in infinitesimal potential wells with nearly infinite central constraint forces, at each event $(\vec{x},t)$ in spacetime.

\section{Conclusions} \label{Conclusions}

The quantum fluid picture, where the quantum potential $Q$ is treated as another classical potential energy in the fluid, provides an intuitive and natural way to think about quantum mechanics.  The total potential $Q+U$ seen by the fluid particles, and their resulting motion, tends to look very similar for superpositions with similar quantum numbers, regardless of the shape of the external potential $U$.

Including the quantum potential provides a natural explanation for tunneling and barrier penetration, because $Q$ tends to cancel out the barrier in $U$ so that the fluid flows through unobstructed.  It also explains the appearance of interference fringes as a series of broad hills and narrow valleys, where the fluid flows quickly across the valleys, and then slows down and accumulates on the hilltops.

The local energy $E_p$ of particles in the fluid also leads to refined notions of when a particle is in a classically forbidden region, and when a particle's kinetic energy is high enough to constitute (soft) superoscillation, and this notion extends Berry's original $Q<0$ bound to general superposition states.  For some purposes, the global band limit $E_+$ is of more interest, so we have also discussed the case of hard superoscillation.

We also explored the above concepts in the alternative $(q_r, k_c)$ picture, where the terms have been reorganized to give a different interpretation, where the fluid contains local symmetric kinetic energy $k_s$ that does not relate to its net flow, and a corresponding reduced quantum potential.  Both the $(q_r, k_c)$ picture, which is probably more physically correct, and the standard $(q,k_a)$ picture, which gives an intuitive explanation of the fluid motion, have their conceptual advantages, so it is useful to consider situations from both perspectives.

We hope that the extensive set of examples provided here will help the reader to build a good physical intuition for the fluid picture and how the quantum potential gives rise to all of the seemingly nonclassical aspects of the fluid dynamics - namely tunneling, interference, and superoscillation.

This work fits into a larger program aimed at developing a fully local interpretation of quantum mechanics in spacetime \cite{waegell2023local, waegell2024toward} (at the cost of having many local worlds defined in one spacetime).  In this model, the standard universal quantum wavefunction is entirely replaced by a collection of fluids in spacetime, with elemental particles in the fluids moving along time/light-like world-lines, and no nonlocal connections of any kind.  All of the physics in this model of quantum mechanics occurs in spacetime, which may be a critical step toward integrating quantum physics and general relativity.

In the shorter term, we aim to develop the relativistic generalization of this particle treatment, for both massive and massless particles, and for both fermions and bosons.  We hope this will resolve the issues with infinite velocity in the nonrelativistic treatment, and that we will obtain a picture where we can understand the Lorentz transform of a generalized quantum potential.

\printbibliography

@article{madelung1927quantum,
  title={Quantum theory in hydrodynamical form},
  author={Madelung, Erwin},
  journal={z. Phys},
  volume={40},
  pages={322},
  year={1927}
}

@article{waegell2023local,
  title={{L}ocal {Q}uantum {T}heory with {F}luids in {S}pace-{T}ime},
  author={Waegell, Mordecai},
  journal={Quantum Reports},
  volume={5},
  number={1},
  pages={156--185},
  year={2023},
  publisher={MDPI}
}

@article{bohm1952suggested,
  title={A suggested interpretation of the quantum theory in terms of ``hidden'' variables. I},
  author={Bohm, David},
  journal={Physical review},
  volume={85},
  number={2},
  pages={166},
  year={1952},
  publisher={APS}
}

@article{silverman1982relativistic,
  title={Relativistic time dilatation of bound muons and the {L}orentz invariance of charge},
  author={Silverman, Mark P},
  journal={American Journal of Physics},
  volume={50},
  number={3},
  pages={251--254},
  year={1982},
  publisher={American Association of Physics Teachers}
}

@article{schonberg1954hydrodynamical,
  title={On the hydrodynamical model of the quantum mechanics},
  author={Sch{\"o}nberg, M},
  journal={Il Nuovo Cimento (1943-1954)},
  volume={12},
  pages={103--133},
  year={1954},
  publisher={Springer}
}

@article{reddiger2017madelung,
  title={The {M}adelung picture as a foundation of geometric quantum theory},
  author={Reddiger, Maik},
  journal={Foundations of Physics},
  volume={47},
  number={10},
  pages={1317--1367},
  year={2017},
  publisher={Springer}
}

@article{bohm1954model,
  title={Model of the causal interpretation of quantum theory in terms of a fluid with irregular fluctuations},
  author={Bohm, David and Vigier, Jean-Pierre},
  journal={Physical Review},
  volume={96},
  number={1},
  pages={208},
  year={1954},
  publisher={APS}
}

@article{reddiger2023towards,
  title={Towards a mathematical theory of the {M}adelung equations: {T}akabayasi’s quantization condition, quantum quasi-irrotationality, weak formulations, and the {W}allstrom phenomenon},
  author={Reddiger, Maik and Poirier, Bill},
  journal={Journal of Physics A: Mathematical and Theoretical},
  volume={56},
  number={19},
  pages={193001},
  year={2023},
  publisher={IOP Publishing}
}

@book{wyatt2005quantum,
  title={Quantum dynamics with trajectories: introduction to quantum hydrodynamics},
  author={Wyatt, Robert E},
  volume={28},
  year={2005},
  publisher={Springer Science \& Business Media}
}

@article{berry2020superoscillations,
  title={Superoscillations and the quantum potential},
  author={Berry, Michael V},
  journal={European Journal of Physics},
  volume={42},
  number={1},
  pages={015401},
  year={2020},
  publisher={IOP Publishing}
}

@article{berry2019roadmap,
  title={Roadmap on superoscillations},
  author={Berry, Michael and Zheludev, Nikolay and Aharonov, Yakir and Colombo, Fabrizio and Sabadini, Irene and Struppa, Daniele C and Tollaksen, Jeff and Rogers, Edward TF and Qin, Fei and Hong, Minghui and others},
  journal={Journal of Optics},
  volume={21},
  number={5},
  pages={053002},
  year={2019},
  publisher={IOP Publishing}
}

@book{aharonov2017mathematics,
  title={The mathematics of superoscillations},
  author={Aharonov, Yakir and Colombo, Fabrizio and Sabadini, Irene and Struppa, Daniele and Tollaksen, Jeff},
  volume={247},
  number={1174},
  year={2017},
  publisher={American Mathematical Society}
}

@article{aharonov2011some,
  title={Some mathematical properties of superoscillations},
  author={Aharonov, Y and Colombo, FABRIZIO and Sabadini, IRENE and Struppa, DC and Tollaksen, J},
  journal={Journal of Physics A: Mathematical and Theoretical},
  volume={44},
  number={36},
  pages={365304},
  year={2011},
  publisher={IOP Publishing}
}

@article{kempf2018four,
  title={Four aspects of superoscillations},
  author={Kempf, Achim},
  journal={Quantum Studies: Mathematics and Foundations},
  volume={5},
  number={3},
  pages={477--484},
  year={2018},
  publisher={Springer}
}

@article{ferreira2006superoscillations,
  title={Superoscillations: faster than the {N}yquist rate},
  author={Ferreira, Paulo Jorge SG and Kempf, Achim},
  journal={IEEE transactions on signal processing},
  volume={54},
  number={10},
  pages={3732--3740},
  year={2006},
  publisher={IEEE}
}

@article{berry2006evolution,
  title={Evolution of quantum superoscillations and optical superresolution without evanescent waves},
  author={Berry, MV and Popescu, Sandu},
  journal={Journal of Physics A: Mathematical and General},
  volume={39},
  number={22},
  pages={6965},
  year={2006},
  publisher={IOP Publishing}
}

@article{berry2008natural,
  title={Natural superoscillations in monochromatic waves in {D} dimensions},
  author={Berry, MV and Dennis, MR},
  journal={Journal of Physics A: Mathematical and Theoretical},
  volume={42},
  number={2},
  pages={022003},
  year={2008},
  publisher={IOP Publishing}
}

@article{aharonov2022unified,
  title={A unified approach to {S}chr{\"o}dinger evolution of superoscillations and supershifts},
  author={Aharonov, Yakir and Behrndt, Jussi and Colombo, Fabrizio and Schlosser, Peter},
  journal={Journal of Evolution Equations},
  volume={22},
  number={1},
  pages={26},
  year={2022},
  publisher={Springer}
}

@article{aharonov2002superoscillations,
  title={Superoscillations and tunneling times},
  author={Aharonov, Yakir and Erez, Noam and Reznik, Benni},
  journal={Physical Review A},
  volume={65},
  number={5},
  pages={052124},
  year={2002},
  publisher={APS}
}

@article{aharonov2010time,
  title={A time-symmetric formulation of quantum mechanics},
  author={Aharonov, Yakir and Popescu, Sandu and Tollaksen, Jeff},
  journal={Physics today},
  volume={63},
  number={11},
  pages={27--32},
  year={2010},
  publisher={AIP Publishing}
}

@article{aharonov2024instability,
  title={Instability and quantization in quantum hydrodynamics},
  author={Aharonov, Yakir and Shushi, Tomer},
  journal={Modern Physics Letters B},
  pages={2450268},
  year={2024},
  publisher={World Scientific}
}

@article{amit2023countering,
  title={Countering a fundamental law of attraction with quantum wave-packet engineering},
  author={Amit, Gal and Japha, Yonathan and Shushi, Tomer and Folman, Ron and Cohen, Eliahu},
  journal={Physical Review Research},
  volume={5},
  number={1},
  pages={013150},
  year={2023},
  publisher={APS}
}

@article{shushi2023reduced,
  title={Reduced role of the wavefunctions' curvature of quantum potentials in non-standard quantum systems},
  author={Shushi, Tomer},
  journal={Physics Letters A},
  volume={475},
  pages={128850},
  year={2023},
  publisher={Elsevier}
}

@article{shushi2023classicality,
  title={Classicality of single quantum particles in curved spacetime through the hydrodynamical reformulation of quantum mechanics},
  author={Shushi, Tomer},
  journal={Journal of Physics A: Mathematical and Theoretical},
  volume={56},
  number={36},
  pages={365301},
  year={2023},
  publisher={IOP Publishing}
}

@article{berry2023time,
  title={Time-independent, paraxial and time-dependent {M}adelung trajectories near zeros},
  author={Berry, Michael},
  journal={Journal of Physics A: Mathematical and Theoretical},
  volume={57},
  number={2},
  pages={025201},
  year={2023},
  publisher={IOP Publishing}
}

@article{berry2023quantum,
  title={Quantum curl forces},
  author={Berry, MV and Shukla, Pragya},
  journal={Journal of Physics A: Mathematical and Theoretical},
  volume={56},
  number={48},
  pages={485206},
  year={2023},
  publisher={IOP Publishing}
}

@article{sanz2019bohm,
  title={Bohm’s approach to quantum mechanics: {A}lternative theory or practical picture?},
  author={Sanz, AS},
  journal={Frontiers of Physics},
  volume={14},
  pages={1--15},
  year={2019},
  publisher={Springer}
}

@article{silva2023properties,
  title={Properties of the Airy beam by means of the quantum potential approach},
  author={Silva-Ortigoza, Gilberto and Ortiz-Flores, Jessica},
  journal={Physica Scripta},
  volume={98},
  number={8},
  pages={085106},
  year={2023},
  publisher={IOP Publishing}
}

@book{holland1995quantum,
  title={The quantum theory of motion: an account of the de {B}roglie-{B}ohm causal interpretation of quantum mechanics},
  author={Holland, Peter R},
  year={1995},
  publisher={Cambridge university press}
}

@article{takabayasi1955vector,
  title={The vector representation of spinning particle in the quantum theory, I},
  author={Takabayasi, Takehiko},
  journal={Progress of Theoretical Physics},
  volume={14},
  number={4},
  pages={283--302},
  year={1955},
  publisher={Oxford University Press}
}

@article{takabayasi1983vortex,
  title={Vortex, spin and triad for quantum mechanics of spinning particle. I: general theory},
  author={Takabayasi, Takehiko},
  journal={Progress of theoretical physics},
  volume={70},
  number={1},
  pages={1--17},
  year={1983},
  publisher={Oxford University Press}
}

@article{takabayasi1983hydrodynamical,
  title={Hydrodynamical formalism of quantum mechanics and Aharonov-Bohm effect},
  author={Takabayasi, Takehiko},
  journal={Progress of Theoretical Physics},
  volume={69},
  number={5},
  pages={1323--1344},
  year={1983},
  publisher={Oxford University Press}
}

@article{takabayasi1954hydrodynamical,
  title={On the hydrodynamical representation of non-relativistic spinor equation},
  author={Takabayasi, Takehiko},
  journal={Progress of Theoretical Physics},
  volume={12},
  number={6},
  pages={810--812},
  year={1954},
  publisher={Oxford University Press}
}

@article{hirschfelder1974quantized,
  title={Quantized vortices around wavefunction nodes. {II}},
  author={Hirschfelder, Joseph O and Goebel, Charles J and Bruch, Ludwig W},
  journal={The Journal of Chemical Physics},
  volume={61},
  number={12},
  pages={5456--5459},
  year={1974},
  publisher={American Institute of Physics}
}

@article{hirschfelder1977angular,
  title={The angular momentum, creation, and significance of quantized vortices},
  author={Hirschfelder, Joseph O},
  journal={The Journal of Chemical Physics},
  volume={67},
  number={12},
  pages={5477--5483},
  year={1977},
  publisher={American Institute of Physics}
}

@article{wu1993quantum,
  title={Quantum probability flow patterns},
  author={Wu, Hua and Sprung, DWL},
  journal={Physics Letters A},
  volume={183},
  number={5-6},
  pages={413--417},
  year={1993},
  publisher={Elsevier}
}

@article{waegell2024toward,
  title={Toward local {M}adelung mechanics in spacetime},
  author={Waegell, Mordecai},
  journal={Quantum Studies: Mathematics and Foundations},
  pages={1--18},
  year={2024},
  publisher={Springer}
}

@article{wu1994inverse,
  title={Inverse-square potential and the quantum vortex},
  author={Wu, Hua and Sprung, DWL},
  journal={Physical Review A},
  volume={49},
  number={6},
  pages={4305},
  year={1994},
  publisher={APS}
}

@article{bloch2023spacetime,
  title={Spacetime superoscillations and the relativistic quantum potential},
  author={Bloch, Yakov},
  journal={Foundations of Physics},
  volume={53},
  number={2},
  pages={46},
  year={2023},
  publisher={Springer}
}

@article{berry2021semiclassical,
  title={Semiclassical superoscillations: interference, evanescence, post-{WKB}},
  author={Berry, MV and Shukla, Pragya},
  journal={New Journal of Physics},
  volume={23},
  number={11},
  pages={113014},
  year={2021},
  publisher={IOP Publishing}
}

@article{nye1974dislocations,
  title={Dislocations in wave trains},
  author={Nye, John Frederick and Berry, Michael Victor},
  journal={Proceedings of the Royal Society of London. A. Mathematical and Physical Sciences},
  volume={336},
  number={1605},
  pages={165--190},
  year={1974},
  publisher={The Royal Society London}
}

@article{barnett2013superweak,
  title={Superweak momentum transfer near optical vortices},
  author={Barnett, Stephen M and Berry, MV},
  journal={Journal of Optics},
  volume={15},
  number={12},
  pages={125701},
  year={2013},
  publisher={IOP Publishing}
}

 \end{document}